\documentclass[12pt]{article}
\usepackage{pstricks,pst-coil}
\usepackage{amsfonts}

\newcommand{\vev}[1]{\left\langle #1\right\rangle}
\newcommand{\bbbz}{{\mathbb Z}}

\newcommand{\bbbp}{{\mathbb P}}
\newcommand{\bbbc}{{\mathbb C}}
\newcommand{\comp}{\bbbc}

\newcommand{\bbbf}{{\mathbb F}}
\newcommand{\bbbr}{{\mathbb R}}

\newcommand{\kcs}{k_{\rm cs}}
\newcommand{\kwz}{k_{\rm wz}}
\newcommand{\wzf}{\Omega_{\rm WZ}}
\newcommand{\wcs}{\Omega_{\rm CS}}

\newcommand{\tr}{\mathop\mathrm{tr}\nolimits}

\newcommand{\Det}{\mathop\mathrm{Det}\nolimits}
\renewcommand{\Im}{\mathop\mathrm{Im}\nolimits}
\renewcommand{\Re}{\mathop\mathrm{Re}\nolimits}
\newcommand{\be}{\begin{equation}}
\newcommand{\ee}{\end{equation}}
\newcommand{\bea}{\begin{eqnarray}}
\newcommand{\eea}{\end{eqnarray}}
\def\fixformat{%
        \normalbaselineskip=20pt plus 0.2pt minus 0.1pt
	\baselineskip=\normalbaselineskip
        \lineskip=2pt plus 0.1pt minus 0.1pt
        \lineskiplimit=2pt
        \abovedisplayskip=15pt plus 5pt minus 3pt
        \belowdisplayskip=\abovedisplayskip
        \parskip=6pt plus 2pt minus 1pt
	\skip\footins=20pt plus 10pt minus 5pt
	\predisplaypenalty=5000
	\postdisplaypenalty=500
	\interlinepenalty=50
	\interdisplaylinepenalty=10000
	\flushbottom
        }
\begin{document}
\fixformat

\begin{titlepage}

\rightline{\vbox{%
    \baselineskip=15pt \tabskip=0pt
    \halign{%
	#\unskip\hfil\cr
	UTTG--14--02\cr
	HUTP--02/A063\cr
	December 9, 2002\cr
	hep-th/0212098 \cr
	}%
    }}
\vskip 2pc plus 1fil

\centerline{\large\bf Quantum Deconstruction of a 5D SYM and its Moduli Space*}
\vskip 3pc plus 1fil
\centerline{\large Amer Iqbal$\strut^{(1)}$ and
	Vadim S.\ Kaplunovsky$\strut^{(2)}$}
\vskip 1pc
\tabskip=0pt plus 3in
\halign to \hsize{%
    \it #\hfil\cr
    \llap{\rm (1)\enspace}\relax
	Jefferson Laboratory\cr
    Harvard University,\cr
    Cambridge, MA~02138, USA\cr
    \tt iqbal@fresnel.harvard.edu\cr
    \noalign{\vskip 1pc}
    \llap{\rm (2)\enspace}\relax
	Theory Group, Department of Physics\cr
    University of Texas at Austin,\cr
    Austin, TX~78712, USA\cr
    \tt vadim@physics.utexas.edu\cr
    }
\tabskip=0pt
\vskip 3pc plus 1fil

\begin{abstract}
We deconstruct the fifth dimension of the 5D SYM theory with $SU(M)$ gauge symmetry
and Chern--Simons level~$\kcs=M$ and
show how the 5D moduli space follows from the non-perturbative analysis
of the 4D quiver theory.
The 5D coupling $h=1/g^2_5$ of the un-broken $SU(M)$ is allowed to take
any non-negative values, but it cannot be continued to $h<0$ and there are
no transitions to other phases of the theory.
The alternative UV completions of the same 5D SYM ---
via M theory on the $\comp^3/\bbbz_{2M}$ orbifold or via the dual five-brane web
in type~IIB string theory --- have identical moduli spaces: $h\ge0$ only,
and no flop transitions.
We claim these are intrinsic properties of the $SU(M)$ SYM theory with
$\kcs=M$. 
\end{abstract}
\vskip 2pc plus 1fil

\begingroup
    \catcode`\@=11
    \let\@thefnmark=\relax
    \@footnotetext{%
	\nobreak
	\par\noindent \hangafter=1 \hangindent=\parindent *\enspace
	Research supported by the US National Science Foundation
	(grants NSF--DMS/00--74329 and PHY--0071512),
	by the Robert~A.\ Welsh foundation, and by the
	US--Israel Bi--National Science Foundation.%
	}
\endgroup

\end{titlepage}
\newpage
\pagenumbering{arabic}

\section{Introduction}
In the course of developing the string/M theory over the last generation, we have
learned much about the ordinary quantum field theories in various spacetime dimensions.
This is particularly true in $D>4$ where interactive field theories are
non-renormalizable and therefore ill-defined at the quantum level --- but embedding into
a string or M theory provides an ultraviolet completion which respects all the important
symmetries of the QFT such as SUSY, gauge and Lorentz invariance \cite{KLMVW,witten97, KKV, 
seiberg, witten96}.
For example, a 5D gauge theory with ${\cal N}=1$ SUSY can be UV completed as:
(1) type~$\rm I^\prime$ string theory on D4--brane probes in a background
    of D8--branes and O8 orientifold planes~\cite{seiberg}, or
(2) M theory compactified on a singular Calabi--Yau threefolds~\cite{DKV,MS,IMS}, or
(2) type~IIB string theory on a web of five-branes~\cite{AHK,LV, KR2}.
With the help of 8 unbroken supercharges, these UV completions allow us to calculate
the exact moduli-dependent gauge couplings of the 5D theory and the global geometry of
its moduli space, including `flop' transitions to different 5D phases \cite{PhasesW}, 
sometimes involving strongly-coupled sectors with non-trivial IR fixed points.

Unfortunately, using a `stringy' UV completion of a 5D supersymmetric gauge theory
to derive its phase structure poses a troubling question:
{\em Is this phase structure an inherent property of the 5D theory
({\sl regardless of a UV regulator}), or is it an artifact of string/M theory?\/}
To resolve this quandary, we need to compare the phase structures of different
UV completions of the same 5D theory.
However, there is no use in comparing the stringy UV completions to each other:
They are dual as string/M theories, and the agreement between their 5D phase
structures  does not prove anything besides confirming the duality.
Instead, the stringy completions should be compared to a completely different,
non-stringy UV completion of the same 5D theory.
And of course, to study the 5D phase structures, we need this non-stringy
completion to work even when the 5D couplings are strong \cite{PhasesW}.

In this article, we use {\em dimensional deconstruction} \cite{ACG} of the fifth
spacetime dimension as a non-stringy UV regulator of a {\em quantum} 5D SYM theory.
That is, we replace the continuous $x^4$ coordinate with a discrete lattice
and re-interpret the 5D gauge symmetry $SU(M)$ as a 4D gauge symmetry $[SU(M)]^N$ ---
one $SU(M)$ factor for each 4D layer $x^4=\ell a$ ($\ell=1,2,\ldots,N$).
In the process, we break 4 out of 8 supersymmetries of the 5D theory
and reduce its Lorentz symmetry down to $SO(1,3)$,
but the broken symmetries re-appear in the continuum limit
of distances much larger than the lattice spacing $a$.
In the opposite limit of short distances, the deconstructed theory is effectively
four-dimensional, renormalizable and asymptotically free;
this makes deconstruction a good UV regulator even when the low-energy couplings
are strong.
Furthermore, at all energies the deconstructed theory has 4 unbroken supercharges,
which allows for exact calculations of holomorphic quantities, such as moduli-dependent
abelian gauge couplings of the Coulomb branch of the theory.

Our main results are (1) the method for converting the exact non-perturbative features
of the deconstructed theories into the exact moduli spaces and phase structures of
the 5D continuum limit, and (2) the fact that these moduli spaces and phase structures
are exactly as for the stringy UV completions of the same 5D theory.
Consequently, we believe that the phase structure is inherent in the 5D theory and
does not depends on the manner of its UV completion.

For simplicity, the analysis of this article is limited to the 5D SYM theories with
$SU(M)$ gauge symmetries and no hypermultiplets
({\it i.~e.}, the flavorless $\rm SQCD_5$).
Furthermore, we use the simplest 4D quiver to implement the latticized fifth dimension,
namely a single $SU(M)_\ell$ vector multiplet for each lattice node $\ell$ and a single
bi-fundamental chiral multiplet $({\bf M}_\ell,\overline{\bf M}_{\ell+1})$ for each
lattice link, without any additional 4D fields.
Consequently, in  the 5D continuum limit, the SYM theory has a specific Chern--Simons
level, namely $\kcs=M$.
Changing the 5D Chern--Simons level or/and adding the flavors involves more complicated
quivers, and we shall present them in a separate follow-up article~\cite{IKtwo}.

But before we delve into the details of deconstruction, {\it etc.}, let us briefly
review the basic features of the parameter/moduli spaces of the 5D SYM theories.
First of all, we need to distinguish between the non-dynamical parameters of a 5D theory
and the dynamical moduli of its exactly degenerate vacua.
In 5D, the scalar fields of an ${\cal N}=1$ SYM theory form a single real adjoint
multiplet of the gauge symmetry $G$ and there is no scalar potential.
Consequently, all the vacua of the theory belong to the Coulomb branch where
$G$ is {\sl generically} broken to its Cartan subgroup $U(1)^r$
and the space of dynamical moduli scalars is simply
$\bbbr^r$ divided by the Weyl group of $G$.
For example, $G=SU(2)$ has rank 1 and Weyl symmetry $\bbbz_2$,
hence the dynamical moduli space $\bbbr/\bbbz_2=\bbbr^+$, or in other words
there is a single real modulus $\phi\ge 0$.
More generally, $G=SU(M)$ has rank $r=M-1$ and 
Weyl symmetry $S_M$, hence the $\bbbr^{M-1}/S_M$ moduli space.
Indeed, the dynamical  moduli of the $SU(M)$ theory are the $M$ eigenvalues 
$(\phi_1,\phi_2,\ldots,\phi_M)$ of an hermitian traceless $M\times M$ matrix,
but they are subject to a linear constraint
\be
\phi_1+\phi_2+\cdots+\phi_M = 0
\label{LinCon}
\ee
and the Weyl symmetry of arbitrary permutations of $\phi_i$.
In other words, the moduli space spans sets of $M$ ordered eigenvalues 
satisfying
\be
\phi_1 \le \phi_2 \le \cdots \le \phi_M
\label{Ordering}
\ee
in addition to (\ref{LinCon}). Generically, all the inequalities 
(\ref{Ordering}) are strict and the $SU(M)$ is completely broken
down to $U(1)^{M-1}$; an equality $\phi_i=\phi_{i+1}=\cdots=\phi_{i+k}$ 
signals a surviving non-abelian subgroup $SU(k+1)\subset SU(M)$.

The metric for the dynamical moduli scalars follows from a cubic 
prepotential ${\cal F}(\phi)$, which also governs the gauge couplings
and the Chern--Simons couplings of the gauge fields:
\bea
{\cal L}_{\rm bosonic}\ &=&
\frac{1}{8\pi^2}\,\frac{\partial^2\,{\cal F}}{\partial\phi_i\,\partial\phi_j}\,\biggl[
	\textstyle{1\over 2}\,D_\mu\phi^i\,D^\mu\phi^j\,
	+\,\textstyle{1\over4}\,F^i_{\mu\nu}\,F^{i,\mu\nu}\,\biggr]
    \nonumber \\
&&{}+\ \frac{1}{96\pi^2}\,\frac{\partial^3\,{\cal F}}
		{\partial\phi_i\,\partial\phi_j\,\partial\phi_k}\,
	\epsilon^{\alpha\beta\gamma\mu\nu}\left[
	A^i_\alpha F^j_{\beta\gamma} F^k_{\mu\nu}\,
	+\,{\hbox{\small non-abelian terms,}\atop\hbox{\small if appropriate}}\right] .
\label{FiveDGauge}
\eea
The Chern--Simons couplings have quantized coefficients,
and for any gauge-invariant UV completion
of the theory these coefficients are completely determined at the
one-loop level of the perturbation theory; there are no higher-order
perturbative or non-perturbative corrections.
Consequently, the prepotential and the gauge couplings are also
completely determined at the one-loop level~\cite{seiberg}.
For the Coulomb branch of the flavorless $SU(M)$, we have
\be
{\cal F}^{SU(M)}(\phi)\
=\ \frac{h}{2}\sum_i\phi_i^2\ +\ \frac{k_{CS}}{6}\sum_i\phi_i^3\
+\ \frac{1}{6}\sum_{i>j}\left|\phi_i-\phi_j\right|^3,
\label{SUMPrepotential}
\ee
and hence
\be
\left[\frac{8\pi^2}{g^2}\right]_{ij}\,
=\, \frac{\partial^{2} {\cal F}^{SU(M)}}{\partial \phi_{i}\,\partial \phi_{j}}\,
=\, \left[ h\,+\,k_{CS}\phi_i\,+\sum_k|\phi_i-\phi_k| \right]\times\delta_{ij}\
-\ \left|\phi_i-\phi_j\right| .
\label{SUMGaugeCouplings}
\ee
%
%
Here $h$ is the tree-level inverse gauge coupling of the unbroken $SU(M)$ theory
and $\kcs$ is its tree-level Chern--Simons coefficient.
{}From the $SU(M)$ point of view, $h$ is a non-dynamical constant parameter,
but it can be promoted to a full vector supermultiplet of a larger theory.
Consequently, we have a combined parameter/moduli space spanned by the $h$
and all the $\phi_i$,
and {\em it is the geometry of this combined parameter/moduli space which is
the real question here}.
Specifically, we want to know {\sl whether $h$ can take negative values}, and if yes,
we want to learn everything about the $h<0$ domain of the parameter/moduli space,
such as the allowed ranges of the $\phi_i$ moduli for $h<0$, the details of the flop
transition between the $h>0$ and $h<0$ domains, and most importantly,
{\sl the physics of the $h<0$ phase of the theory}.

The answers to these questions depend on the Chern--Simons level of the SYM theory
in question.
The theory with $\kcs=M$ (or equivalently $\kcs=-M$) --- the main subject of this
article --- does not have any $h<0$ phases;
instead, the parameter space is limited to $h\ge 0$ and there are no flop transitions.
On the other hand, for $|\kcs|<M$ the parameter space has both $h>0$ and $h<0$
domains connected via a flop transition at the infinite-coupling point $h=0$.
Physically, the $h<0$ phase has a not-trivial IR fixed point and its low-energy
limit is governed by a 5D superconformal field theory rather than SYM.

Deriving these phase structures involves non-perturbative analysis of the
strongly coupled 5D SYM --- or rather of its UV completion.
We claim however that different UV completions of the same 5D SYM have identical
phase structures, which are therefore inherent properties of the 5D theory itself,
regardless of the completion.
In this article, we verify this claim for the $\kcs=M$ theory by comparing its
UV completions via M/string theory with the completion via dimensional deconstruction.
The theories with $|\kcs|<M$ will be addressed in detail in the follow-up
article~\cite{IKtwo}.

The rest of this article is organized as follows:
In the following section~2, we study the stringy UV completions of the 5D SYM theories.
In~\S2.1, we put M~theory on a CY orbifold singularity $\comp^3/\bbbz_{2M}$ and show
that a partial resolution of this singularity gives rise to a flavorless $SU(M)$ in 5D.
The moduli/parameter space of the 5D Coulomb branch follows from the intersections
of the 4--cycles of the completely resolved singularity, and we show that this
orbifold model has no flop transitions and that its prepotential corresponds
to $\kcs=M$.
In~\S2.2 we construct a dual type~IIB string model in which the 5D physics
arises from a web of $(p,q)$ five-branes \cite{AHK};
the $SU(M)$ gauge symmetry follows from $M$ parallel five-branes of the same type.
For $\kcs=\pm M$, the web has a unique topology with 2 parallel external legs
separated by the distance $h$; consequently, the 5D SYM has no flop transitions
and only the $h\ge 0$ values are allowed.
But for $|\kcs|<M$, none of the external legs are parallel, and their relative motion
allows both positive and negative values of the $h$ parameter.

In section~3 we focus on the dimensional deconstruction.
After a brief review of deconstruction at the semi-classical level,
we consider the quantum effects in the 4D $[SU(M)]^N$ quiver theory.
In~\S3.1, we use holomorphy (due to 4 unbroken supercharges of the
deconstructed theory) to construct the  moduli space
of the Coulomb branch of the quantum quiver and the Seiberg--Witten curve encoding
the moduli dependence of the gauge couplings.
In the large quiver limit $N\to \infty$ we recover the un-compactified 5D theory,
and we show that the moduli dependence of its gauge couplings is exactly
as in eq.~(\ref{SUMGaugeCouplings}) for $\kcs=M$.
Furthermore, the quantum corrections to the quiver's moduli space translate
into 5D terms as $h\ge 0$, and indeed there are no flop transitions in
the large $N$ limit of this quiver.
In~\S3.2, we deconstruct the 5D Chern--Simons coupling of the un-broken $SU(M)$
theory directly from the quiver, without any help from SUSY.
We show that the quiver theory has Wess--Zumino couplings with coefficients $\kwz=M$
for every bi-linear scalar field of the quiver, which deconstruct the 5D
Chern--Simons couplings at the level $\kcs=\kwz=M$.
This confirms the 5D identity of the deconstructed theory, and it also tells us
we need more complicated quivers to deconstruct the 5D theories with $\kcs\neq\pm M$.

Finally, in the Appendix we calculate the abelian gauge couplings from
the hyperelliptic Seiberg--Witten curve of the 4D quiver constructed in~\S3.1.

\newpage
\section{M Theory Orbifolds and $(p,q)$ Five--Brane Webs}
In this section we discuss the string/M--theory based UV completions
of the 5D SYM theories.
First, we compactify M Theory on singular Calabi--Yau threefolds
and show that a particular $\comp^3/\bbbz_{2M}$ orbifold singularity
(or rather its partial blowup) gives rise to the flavorless $SU(M)$ SYM
in five dimensions \cite{PhasesW}.
The parameter space of this model had no flop transitions, and the
Chern--Simons level of this model turns out to be $\kcs=M$.

In the second half of this section, we take a dual IIB string theory
embedding in which the 5D physics arises from a web of $(p,q)$ five-branes 
\cite{AHK}.
The web diagram dual to the $\comp^3/\bbbz_{2M}$ orbifold has $M$
parallel internal lines --- which explains the $SU(M)$ gauge symmetry
--- and two parallel external legs
--- which explains the $\kcs=M$ and the absence of flop transitions.

\subsection{M Theory on the $\comp^3/\bbbz_{2M}$ Orbifold}
M theory compactified on a {\em smooth} Calabi--Yau threefold gives rise
to an abelian gauge theory in the five un-compactified dimensions.
The 5D gauge fields $A^i_\mu$ follow from the 11D 3--form $C_{\lambda\mu\nu}$
reduced on the $H^{(1,1)}$ cohomology of the threefold,
\be
C\ =\sum_{i=1}^{h_{1,1}} A^i\wedge \omega^i,\qquad
\omega^i\in H^{(1,1)}(\mathrm{CY}) .
\ee
To get a non-abelian 5D gauge symmetry we need a singularity in which
a complex surface $\mathbb S$ collapses to a complex curve $B$ \cite{PhasesW, IMS}.
In this collapse, the 2--cycles contained in $\mathbb S$ and fibered
over the base $B$ shrink to zero area,
and the M2-branes and anti-M2-branes wrapped around these shrinking cycles
give rise to the massless non-abelian vector multiplets.
The rank $r$ of the non-abelian symmetry is given by the number of
homologically-independent 4--cycles $(S_1,\ldots,S_r)$ involved in the
collapsing surface $\mathbb S$
while the Dynkin diagram depends on the intersection matrix of the shrinking
2--cycles fibered over the whole base.
The exceptional fibers --- if any ---
give rise to the matter fields, {\it i.~e.}\ the charged hypermultiplets 
\cite{IMS}.

The inverse gauge coupling $h=(8\pi^2/g^2)$ of the 5D theory
is proportional to the area of the base cycle~$B$.
When the base area shrinks to zero as well, the theory becomes infinitely strongly
coupled and its infrared limit becomes superconformal.
For some singularities, the base cannot be un-shrunk apart from the fibers
and the 4--cycle family $(S_1,\ldots,S_r)$ collapse to a single point;
such singularities give rise to exotic 5D superconformal theories which do not
follow from $g=\infty$ gauge theories.
The best known example of such exotic SCFT is the $E_0$ of rank $r=1$ 
\cite{MS,DKV}.

The simplest example of a singularity where a complex surface collapses
to a point is an orbifold point $\comp^3/\bbbz_N$.
The orbifolding symmetry $\bbbz_N$ acts according to
\be
\bbbz_N:\ (z_1,z_2,z_3)\ \mapsto\
(e^{2\pi i a/N}z_1, e^{2\pi i b/N}z_2, e^{2\pi ic/N}z_3),\qquad
a+b+c=N,
\ee
and the fixed point $(0,0,0)$ hides a blown-down weighted projective space
$\mathbb{WP}^2$ with weights proportional $(a,b,c)$ modulo $N$.
Often, this $\mathbb{WP}^2$ is itself singular, hence a complete resolution
involves $r>1$ homologically-independent 4--cycles, and when these cycles
re-collapse to a line rather than a point, one obtains a 5D gauge theory
of a non-abelian rank $r>1$.

Without loss of generality we assume $a=1$, hence the collapsed surface is
\be
\mathbb S\ =\ \mathbb{WP}^2[1,b,c]
\label{WPspace}
\ee
whose toric diagram  is a simple triangle
with vertices $(1,0)$, $(0,1)$ and $(-c,-b)$;
for example, the diagram for $\mathbb{WP}^2[1,4,7]$ is 
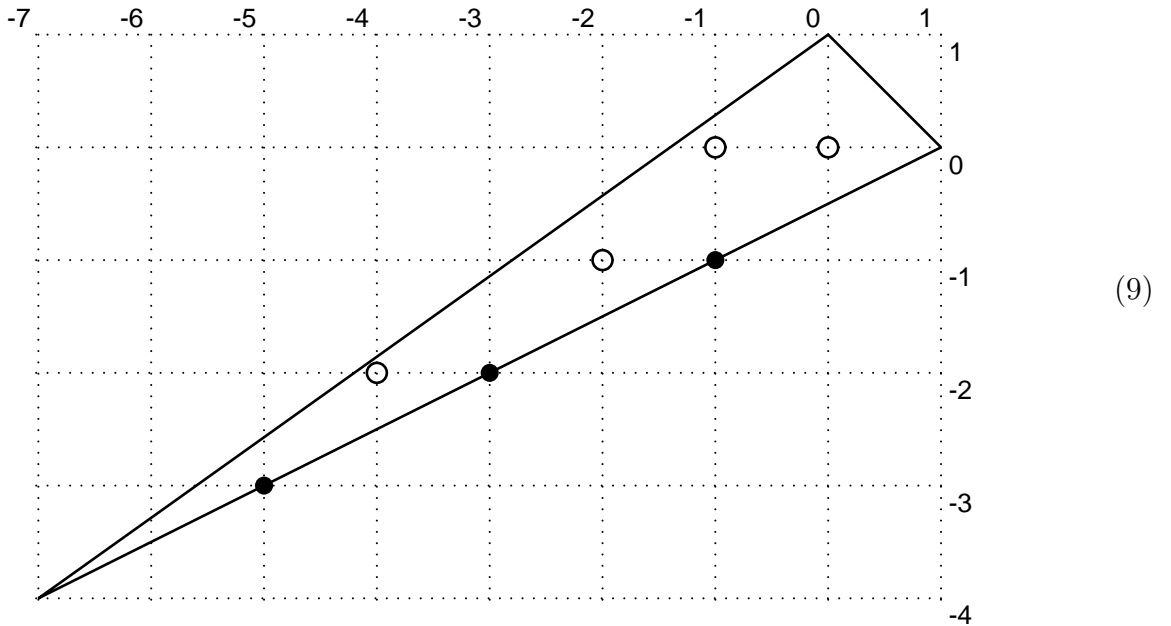
\bea
\psset{unit=15mm,linewidth=1pt,linecolor=black}
\begin{pspicture}[0.5](-7,-4)(+1.3,+1.3)
\psgrid[subgriddiv=1,griddots=10](1,1)(-7,-4)
\psline(1,0)(0,1)(-7,-4)(1,0)
\pscircle(0,0){.1}
\pscircle(-1,0){.1}
\pscircle(-2,-1){.1}
\pscircle(-4,-2){.1}
\pscircle*(-1,-1){.08}
\pscircle*(-3,-2){.08}
\pscircle*(-5,-3){.08}
\end{pspicture}
\label{ToricExample}
\eea
The black circles on this diagram denote integer grid points crossed by the
triangle's side; they correspond to the blow-up modes of the singular line
$z_1=z_3=0$ fixed by the $\bbbz_4\subset \bbbz_{12}$.
Physically, such modes are the non-dynamical parameters of the 5D gauge theory
such as the inverse gauge coupling $h$ or the masses of hypermultiplets.
The open circles denote the grid points completely inside the triangle, which
correspond to the homologically-independent
compact 4--cycles involved in resolving the fixed point itself.
Physically, they become the dynamical moduli vector-multiplets of the Coulomb
branch of the 5D theory and their number is the rank $r$ of the gauge symmetry.

We are looking for the $SU(M)$ SYM theory which has rank $r=M-1$ and a single
non-dynamical parameter $h$,
hence we need a singularity whose resolution has $M-1$ compact 4--cycles,
plus one non-compact cycle responsible for the $h$ parameter.
The simplest orbifold of this kind has $a=b=1$ and $N=2M$,
\be
\bbbz_{2M}:\ (z_1,z_2,z_3)\ \mapsto\
(e^{+2\pi i/2M} z_1, e^{+2\pi i/2M}z_2, e^{-4\pi i/2M}z_3),
\label{ZtwoM}
\ee
and toric diagram of its blowup is $SL(2,\bbbz)$ equivalent to
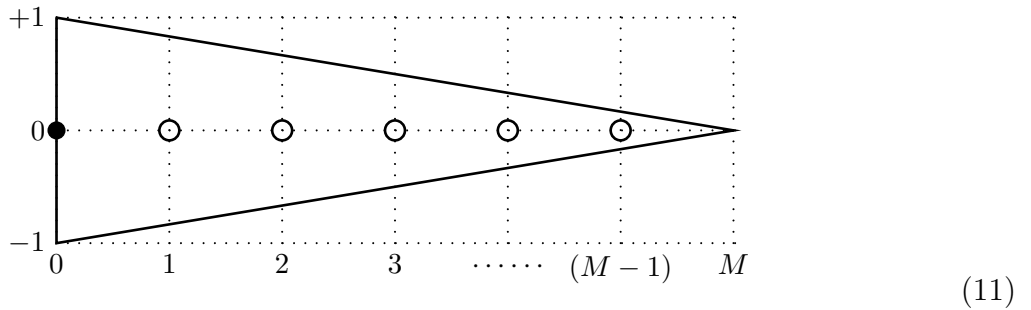
\be
\psset{unit=15mm,linewidth=1pt,linecolor=black}
\begin{pspicture}(-0.5,-1.5)(+6.5,+1.5)
\psgrid[subgriddiv=1,griddots=10,gridlabels=0](0,-1)(+6,+1)
\rput[r](-.1,+1){\small$+1$}
\rput[r](-.1,0){\small$0$}
\rput[r](-.1,-1){\small$-1$}
\rput[t](0,-1.1){\small$0$}
\rput[t](1,-1.1){\small$1$}
\rput[t](2,-1.1){\small$2$}
\rput[t](3,-1.1){\small$3$}
\rput[t](4,-1.1){\small$\vphantom{0}\cdots\cdots{}$}
\rput[t](5,-1.1){\small$(M-1)$}
\rput[t](6,-1.1){\small$M$}
\psline(0,+1)(6,0)(0,-1)(0,+1)
\pscircle*(0,0){0.08}
\pscircle(1,0){.1}
\pscircle(2,0){.1}
\pscircle(3,0){.1}
\pscircle(4,0){.1}
\pscircle(5,0){.1}
\end{pspicture}
\label{ToricSUM}
\ee
It turns out that this orbifold indeed gives rise to the $SU(M)$ SYM in 5D.
Or rather, it gives rise to this SYM after a partial resolution which
blows up the fixed line at $z_1=z_2=0$ and turns up
$h>0$ ({\it i.~e.}\ $g<\infty$) but not the $SU(M)$--breaking moduli fields.
To see how this works, let us orbifold in two stages
\be
\comp^3/\bbbz_{2M}[1,1,(2M-2)]\
=\ \left( \comp^2/\bbbz_2 \otimes \comp \right) /\bbbz_M
\label{TwoStages}
\ee
and blow up the first stage only.
That is, we blowup the $\comp^2/\bbbz_2$ factor without affecting the complex
$z_3$ plane or resolving the singularities due to the second-stage orbifolding
by the $\bbbz_M$ factor.
In the linear sigma model description \cite{wittenlinear} of this partial blowup,
the $\comp^2/\bbbz_2$ is spanned by 3 complex fields
\bea
(Z_1,Z_2,X_0)\ \mathrm{modulo} &&
U(1):\ Z_1\mapsto e^{+i\alpha}Z_1,\ Z_2\mapsto e^{+i\alpha}Z_2,\ 
        X_0\mapsto e^{-2i\alpha}X_0, \nonumber \\
\mathrm{constrained\ to} &&
|Z_1|^2+|Z_2|^2-2|X_0|^2\ =\ D.
\label{LSMtwo}
\eea
In string theory, the linear sigma model is in the geometric CY (or rather K3)
phase for $D>0$ and in the non-geometric Landau--Ginzburg phase for $D<0$,
but only the geometric phase is allowed in the M theory \cite{PhasesW}, 
hence $D\ge0$ and the parameter space of the blowup is $\bbbr^+$ rather 
than $\bbbr$.

The second-stage orbifold action combines the (\ref{ZtwoM}) with
the $U(1)$ action into
\be
\bbbz_M:\ (X_0,Z_1,Z_2,Z_3)\
\mapsto\ (e^{+2\pi i/M}X_0,Z_1,Z_2,e^{-2\pi i/M}Z_3) ,
\ee
which has a fixed {\sl line} at $X_0=Z_3=0$ spanned by the
$(Z_1,Z_2)\in\bbbp^1$ of area $D>0$.
In other words, the threefold geometry has an $A_{M-1}$ singularity
(of complex codimension 2) fibered over the base $B$ which is rational curve
with normal bundle ${\cal O}(0)\oplus {\cal O}(-2)$.
%
%
In M theory, this singularity gives rise to the $SU(M)$ SYM theory
living in the remaining 7 (real) dimensions
comprising $B\times\bbbr^{4,1}_{\rm Minkowski}$,
and after reducing the two compact dimensions of the base $B$,
we end up with a 5D $SU(M)$ SYM at finite gauge coupling
\be
g_5^2\ \propto\ \frac{1}{\mathop{\rm Area}(B)}\ <\ \infty,\qquad
i.~e.,\quad h\,=\,D\, >\, 0.
\ee

To study the Coulomb branch of the 5D $SU(M)$ in M theory, we need a complete
resolution of the orbifold singularity.
In the linear sigma model language, resolving the second-stage $\bbbz_M$
singularity at $X_0=Z_3=0$ involves $(M-1)$ additional complex fields
and a like number of $U(1)$ symmetries and D--term constraints;
altogether we have $(Z_1,Z_2,Z_3;X_0,X_1,\ldots,X_{M-1})$ modulo
\be
U(1)^M:\cases{
        Z_1\mapsto \exp(i\alpha_0)Z_1,\cr
        Z_2\mapsto \exp(i\alpha_0)Z_2,\cr
        Z_3\mapsto \exp(i\alpha_{M-1})Z_3,\cr
        X_0\mapsto \exp(i\alpha_1-2i\alpha_0)X_0,\cr
        X_1\mapsto \exp(i\alpha_2-2i\alpha_1)X_1,\cr
        X_j\mapsto \exp(i\alpha_{j+1}-2i\alpha_j+i\alpha_{j-1})X_j &
                for $j=2,\ldots,(M-2)$\cr
        X_{M-1}\mapsto \exp(-2i\alpha_{M-1}+i\alpha_{M-2}) X_{M-1},\cr
        }
\label{PhaseShifts}
\ee
and constrained by
\be
\cases{
        |Z_1|^2\,+\,|Z_2|^2\,-\,2|X_0|^2\ =\ D>0,\cr
        |Z_3|^2\,+\,|X_{M-2}|^2\,-\,2|X_{M-1}|^2\ =\ D_{M-1}>0,\cr
        |X_{j-1}|^2\,+\,|X_{j+1}|^2\,-\,2|X_j|^2\ =\ D_j>0 &
                for $j=1,2,\ldots,(M-2)$.\cr
        }
\label{RadialConstraints}
\ee
In terms of the toric diagram (\ref{ToricSUM}), the $X_1,\ldots,X_{M-1}$
are the grid points inside the triangle (open circles), the $X_0$ is the 
grid point on the left side (the black circle) and the $Z_1$, $Z_2$ and
$Z_3$ are the triangle's vertices.

\par\vskip 0pt plus 2in \penalty-1000

In real terms, the non-compact CY threefold described by
eqs.~(\ref{PhaseShifts}--\ref{RadialConstraints}) has three independent
phases and three independent radii $|Z_1|^2$,
$|Z_2|^2$ and $|Z_3|^2$;
the constraints (\ref{RadialConstraints})
impose a concave ``floor'' keeping these radii away from the origin:
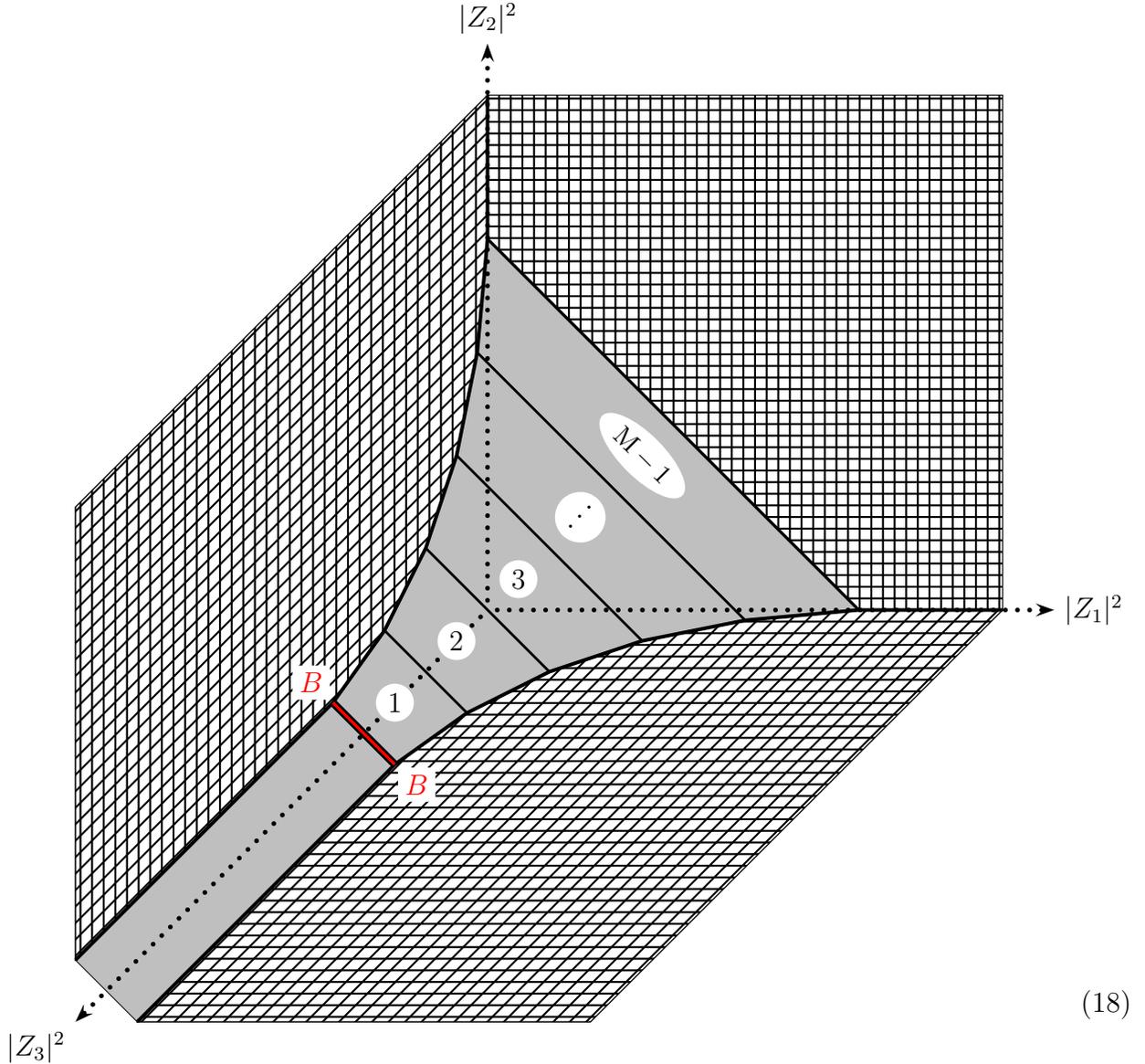
\be
\psset{unit=1.5mm,linewidth=1pt,linecolor=black}
\begin{pspicture}[0.1](-45,-50)(+55,+60)
\pspolygon[linewidth=0,fillstyle=hlines,hatchangle=0]%
    (50,0)(36,0)(25,-1)(15,-3)(6,-6)(-2,-10)(-9,-15)(-34,-40)(10,-40)
\pspolygon[linewidth=0,fillstyle=hlines,hatchangle=45]%
    (50,0)(36,0)(25,-1)(15,-3)(6,-6)(-2,-10)(-9,-15)(-34,-40)(10,-40)
\pspolygon[linewidth=0,fillstyle=hlines,hatchangle=45]%
    (0,50)(0,36)(-1,25)(-3,15)(-6,6)(-10,-2)(-15,-9)(-40,-34)(-40,10)
\pspolygon[linewidth=0,fillstyle=vlines,hatchangle=0]%
    (0,50)(0,36)(-1,25)(-3,15)(-6,6)(-10,-2)(-15,-9)(-40,-34)(-40,10)
\pspolygon[linewidth=0,fillstyle=crosshatch,hatchangle=0]%
    (36,0)(0,36)(0,50)(50,50)(50,0)
\psline[linewidth=1.5pt](36,0)(0,36)
\pspolygon[fillstyle=solid,fillcolor=lightgray](36,0)(0,36)(-1,25)(25,-1)
\pspolygon[fillstyle=solid,fillcolor=lightgray](15,-3)(-3,15)(-1,25)(25,-1)
\pspolygon[fillstyle=solid,fillcolor=lightgray](6,-6)(-6,6)(-3,15)(15,-3)
\pspolygon[fillstyle=solid,fillcolor=lightgray](-2,-10)(-10,-2)(-6,6)(6,-6)
\pspolygon[fillstyle=solid,fillcolor=lightgray](-9,-15)(-15,-9)(-10,-2)(-2,-10)
\pspolygon[fillstyle=solid,fillcolor=lightgray,linewidth=0]%
    (-34,-40)(-40,-34)(-15,-9)(-9,-15)
\psline[linewidth=1.5pt](50,0)(36,0)(25,-1)(15,-3)(6,-6)(-2,-10)(-9,-15)(-34,-40)
\psline[linewidth=1.5pt](0,50)(0,36)(-1,25)(-3,15)(-6,6)(-10,-2)(-15,-9)(-40,-34)
\psline[doubleline=true,doublecolor=red](-15,-9)(-9,-15)
\psline[linestyle=dotted,linewidth=2pt]{->}(0,0)(55,0)
\psline[linestyle=dotted,linewidth=2pt]{->}(0,0)(0,55)
\psline[linestyle=dotted,linewidth=2pt]{->}(0,0)(-40,-40)
\rput*[lt](-8,-16){\red$B$}
\rput*[rb](-16,-8){\red$B$}
\cput*[linecolor=white](-9,-9){1}
\cput*[linecolor=white](-3,-3){2}
\cput*[linecolor=white](+3,+3){3}
\cput*[linecolor=white]{45}(+9,+9){$\cdots$}
\rput{315}(15,15){\small\psovalbox*[linecolor=white]{$\rm M-1$}}
\rput[l](56,0){$|Z_1|^2$}
\rput[b](0,56){$|Z_2|^2$}
\rput[tr](-41,-41){$|Z_3|^2$}
\end{pspicture}
\label{FloorDiagram}
\ee
In this picture, the floor ``tiles'' labeled
 $\pscirclebox{1},\pscirclebox{2},\ldots,\psovalbox{M-1}$
represent the compact 4--cycles $S_1,S_2,\ldots,S_{M-1}$ which collapse to
a common base $B$ when the $SU(M)$ symmetry is restored.
{}From the linear sigma model point of view, it is clear that the base $B$
(the red line in~(\ref{FloorDiagram})) is a $\bbbp^1$ and
each 4--cycle $S_j$ is a $\bbbp^1$ fibration over $B$.
The base area $D$ governs the 5D gauge coupling while the fiber areas $D_j$
control the $SU(M)$ breaking down to the Cartan subgroup $U(1)^{M-1}$;
in 5D terms,
\be
D\ =\ h,\qquad D_j\ =\ \phi_j\,-\,\phi_{j+1} .
\label{PoneAreas}
\ee

Blowing down the fibers but not the base un--Higgses the $SU(M)$, but what happens
when we blow down the base but not the fibers?
Mathematically, this is equivalent to making the fibers very large while the base~$B$
remains finite, and in that limit, the CY geometry becomes
\be
\mathrm{CY}\ \longrightarrow\ \comp\times T^*(\bbbp^1) .
\ee
The $T^*(\bbbp^1)$ does not admit flop transitions to anything else,
which means that the resolved orbifold also does not have any flop transitions
to a physically different phase with $h<0$.\footnote{%
        The resolved orbifold does have flop transitions to negative {\em fiber}
        areas $D_j<0$, but these are just the Weyl symmetries of the $SU(M)$.}
Thus, we conclude that
{\bf the UV completion of the 5D $SU(M)$ SYM theory via M theory on the
$\comp^3/\bbbz_{2M}$ orbifold has parameter space $\bbbr^+$ and no phase
transitions}.

In the following sections, we shall obtain the same parameter space for the
UV completions via 5--brane webs in the type~IIB string theory or via dimensional
deconstruction, {\em provided} the SYM theory has Chern--Simons level $\kcs=M$.
To compare the results, we need the Chern--Simons level of the SYM derived
from the M~theory on the $\comp^3/\bbbz_{2M}$.
We shall see momentarily that this Chern--Simons level is indeed $\kcs=M$,
hence the $\bbbr^+$ parameter space is appears to be the intrinsic property
of the $SU(M)$ SYM theory at this Chern--Simons level regardless of the UV completion.

As a warm-up exercise, consider the $SU(2)$ case.
In 5D, the $SU(2)$ SYM does not have Chern--Simons interactions
(due to lack of a cubic invariant), but thanks to $\pi_4(SU(2))=\bbbz_2$ 
there is a discrete vacuum angle ($\theta=0$ or $\theta=\pi$)
and hence two distinct quantum theories.
Their respective $g\to\infty$ superconformal limits are called the
$E_1$ and the $\tilde E_1$; they belong to the $E$ series of the
rank~1 SCFTs whose global symmetries form the exceptional $E_n$ groups,
hence the name~\cite{DKV,MS}.
In M theory, an $E_n$ SCFT shows up when a del~Pezzo surface ${\cal B}_n$  
collapses to a point \cite{DKV, MS}; the surfaces ${\cal B}_1$ and 
$\widetilde{{\cal B}}_1$ giving rise to the $E_1$ and the $\tilde E_1$ 
happen to be $\bbbp^1$ fibrations over the $\bbbp^1$ base without bad fibers.
In other words, they are Hirzebruch ruled surfaces
\be
\bbbf_n\ \buildrel{\rm def}\over= \
\bbbp\bigl({\cal O}(0)\oplus{\cal O}(n)\bigr) ;
\label{Hirzebruch}
\ee
specifically, ${\cal B}_1=\bbbf_0=\bbbp^1\times\bbbp^1$ and
$\widetilde{{\cal B}}_1=\bbbf_1$.

When a Hirzebruch surface $\bbbf_n$ collapses to a line rather than a point 
because its base $B$ retains finite area while the fiber area shrinks to zero,
the M theory on the resulting singularity produces an $SU(2)$ SYM with
a finite gauge coupling instead of an SCFT.
This works for any fibration degree $n$, but the parity of $n$ affects the
vacuum angle \footnote{As we will see in the next section from the type IIB
web perspective the Hirzebruch surface $\bbbf_{n}$ for $n>2$ must
be accompanied by other 4-cycle inside a CY3-fold.}:
\be
\theta\ =\ n\pi\ \mathrm{modulo}\ 2\pi .
\label{VacuumAngle}
\ee
In the $\comp^3/\bbbz_{2M}$ orbifold model for $M=2$, the compact 4--cycle of the
fully resolved singularity is the blown-up weighted projective space $\mathbb{WP}^2[1,1,2]$.
The blow-up removes the $\bbbz_2$ orbifold singularity of the $\mathbb{WP}^2[1,1,2]$
itself and turns it into the Hirzebruch surface $\bbbf_2$ of degree 2.
For the partially resolved $\bbbz_4$ singularity, the $\bbbf_2$ collapses to its base $B$
and the $SU(2)$ is un--Higgsed.
According to eq.~(\ref{VacuumAngle}), the vacuum angle of this $SU(2)$
is zero rather than~$\pi$.

The orbifold models with $M>2$ have several compact 4--cycles $S_1,\ldots,S_{M-1}$,
{\it cf.}\ the picture~(\ref{FloorDiagram}).
Let us calculate their triple intersections $S_j\cdot S_k\cdot S_\ell$
in the fully resolved CY3-fold.
According to \cite{IMS}, this will give us the prepotential for the $M-1$ abelian vector
multiplets of the Coulomb branch of the 5D SYM:
In terms of the Cartan moduli $a_i=\phi_{i+1}-\phi_i$,
\be
{\cal F}_3\ \equiv \ \mbox{the cubic part of}\ {\cal F}\
=\ \frac{1}{6}\left[\sum_{i,j=1}^{M-1} C^{ij}a_i\,S_j\right]^3
\ee
where $C^{ij}$ is the inverse Cartan matrix of the $SU(M)$, or in terms
of the eigenvalues $\phi_i$,
\be
{\cal F}_3\
=\ \frac{1}{6}\left[\sum_{j=1}^{M-1}(\phi_{j+1}+\cdots+\phi_M)\,S_j\right]^3
\label{CubicF}
\ee
Physically, the result of this calculation should agree with the $SU(M)$ Coulomb
branch prepotential~(\ref{SUMPrepotential}), and we shall see momentarily that
this is indeed the case for $\kcs=M$.

The toric diagram of~(\ref{ToricSUM}) of the orbifold implies that the 4--cycle~$S_j$
is a Hirzebruch surface~$\bbbf_{2j}$.
Indeed, in the linear sigma model language 
({\it cf.}\ eqs.~(\ref{PhaseShifts}--\ref{RadialConstraints}) and the picture
(\ref{FloorDiagram})),  $S_j$ is the locus of $X_j=0$ where sigma model reduces to
just 4 fields $(Z_1,Z_2,X_{j-1},X_{j+1})$ with 2 abelian symmetries and 2 constraints:
\bea
U(1)^2:\ (Z_1,Z_2,X_{j-1},X_{j+1}) &\mapsto&
(e^{i\alpha_0}Z_1,e^{i\alpha_0}Z_2,e^{i(\alpha_j-2j\alpha_0)}X_{j-1},e^{i\alpha_j}X_{j+1}),
        \nonumber\\
|Z_1|^2\,+\,|Z_2|^2\,-\,2j|X_{j-1}|^2 &=&
D\,+\,2D_1\,+\,4D_2\,+\,\cdots\,+\,(2j-2)D_{j-1}\ >\ 0,\nonumber \\
|X_{j-1}|^2\,+\,|X_{j+1}|^2 &=& D_j\ >\ 0, 
\eea
which clearly describes the ruled surface~$\bbbf_{2j}$.
The second homology of this surface is spanned by the fibration base $B_j$ and the fiber $F_j$
which intersect according to
\be
B_j\cdot B_j\ =\ -2j,\qquad B_j\cdot F_j\ =\ 1,\qquad F_j\cdot F_j\ =\ 0
\label{BFIntersect}
\ee
and the canonical class dual to the first Chern class $c_1(S_j=\bbbf_{2j})$ is
\be
-K_{S_j}\ =\ 2B_j\ +\ (2j+2)F_j.
\label{CanonicalClass}
\ee
In the Calabi--Yau context, this canonical class controls the intersections of $S_j$
with lines equivalent to rational curves within the $S_j$ itself:
\be
\forall C\subset S_j:\
\left[S_j\cdot C\right]_{\rm CY}\ =\ \left[K_{S_j}\cdot C\right]_{S_j}
\label{UsingK}
\ee
The curves of interest for our purposes are the intersections of $S_j$ with itself
or with the other $S_i$.
The self-intersection $S_j\cdot S_j$ can be found by deforming
the $S_j$ inside the Calabi--Yau threefold and finding the vanishing locus of the
deformation.
The deformations normal to the $S_j$ are sections of the canonical class $K_{S_j}$
and vanish on the canonical class itself.
Hence, $S_j\cdot S_j= K_{S_j}$ and therefore (thanks to eq.~(\ref{UsingK}))
\be
S_j\cdot S_j\cdot S_j\ =\ K_{S_j}\cdot K_{S_j}\ =\ 8 .
\label{SelfIntersection}
\ee
Besides itself, the cycle $S_j$ intersects its nearest neighbors $S_{j\pm1}$
but not other $S_i$ with $i\neq j,j\pm1$ --- this is obvious from the
diagram~(\ref{FloorDiagram}).
For the neighbors,
\be
S_j\cdot S_{j+1}\ =\ B_{j+1}\ =\ B_j\,+\,2j F_j
\ee
and therefore
\bea
S_j\cdot S_{j+1}\cdot S_{j+1} &=&
B_{j+1}\cdot K_{S_{j+1}}\ =\ +2j ,\label{JNN} \\
S_j\cdot S_{j+1}\cdot S_j &=&
(B_j+2jF_j)\cdot K_{S_j}\ =\ -(2j+2) .\label{JNJ}
\eea
All other triple intersections of the $S_j$ cycles vanish,
hence substituting eqs.~(\ref{SelfIntersection}) and (\ref{JNN}--\ref{JNJ}) into
eq.~(\ref{CubicF}) we arrive (after some messy algebra) at the cubic prepotential
\be
{\cal F}_3\ =\ \frac{1}{12}\sum_{ij}|\phi_i-\phi_j|^3\
+\ \frac{M}{6}\sum_i\phi_i^3 .
\ee
As promised, this prepotential is exactly as in eq.(\ref{SUMPrepotential}) for
$\kcs=M$.
Thus, we identify the low-energy limit of the M~theory on the resolved orbifold as
the Coulomb branch of the $SU(M)$ SYM {\em at the Chern--Simons level $\kcs=M$}.

The SYM theories with other Chern--Simons levels also have UV completions via M
theory compactified on singular CY threefolds, but the singularities for 
$|\kcs|\neq M$ are not orbifolds.
They also have $\bbbr$ rather than $\bbbr^+$ parameter spaces, and they do have
flop transitions to physically different phases for $h<0$ for $M-k$ odd.
In the next section, we shall derive these results from the dual description
in terms of  $(p,q)$ 5--brane webs of the IIB superstring.

\newpage
\subsection{Five--Brane Webs in IIB Superstring Theory}
The type~IIB superstring theory has BPS five-branes with any mutually-prime
combinations $(p,q)$ of the R-R and NS-NS magnetic charges.
A brane web \cite{AHK,KR2} comprises a network of such five-branes
spanning $1+4$ common infinite dimensions $(x^0,x^1,x^2,x^3,x^4)$
while the fifth space dimension of each $(p,q)$ brane is a line
in the common $(x^5,x^6)$ plane ($x^7=x^8=x^9=0$)
in the direction
\be
x^5\,+\,ix^6\ =\ p\,+\,q\tau
\label{Slope}
\ee
where $\tau_{\rm IIB}=\frac{4\pi i}{g^2}+\frac{a}{2\pi}$ is the ``holomorphic''
type~IIB string coupling.
The lines may be finite or infinite in one direction; generally, they split
and join each other at trivalent vertices where eqs.\
\be
\sum_{i=1}^3 p_i\ =\sum_{i=1}^3 q_i\ =\ 0
\label{FiveBraneCharges}
\ee
are required by the charge conservation.
Together, eqs.~(\ref{Slope}) and (\ref{FiveBraneCharges}) assure the mechanical
stability of the five-brane web --- as well as the ${\cal N}=1$ supersymmetry
of the effective 5D theory which lives on it.

The 5D theory living on a generic brane web is an abelian gauge theory coupled
to massive charged particles arising from the open strings (or webs of open strings)
ending on the five-branes.
The rank of this 5D theory is the number of loops in the graph of the brane
web in the 56 plane, and the moduli correspond to normal movements
of the finite-length internal lines in this graph while the semi-infinite
external lines remain fixed.
Moving the external lines of the graph relative to each other corresponds
to changing the non-dynamic parameters of the 5D theory such as $h$ or
quark masses \cite{AHK}.

A movement which puts several parallel lines directly on top of each other
un--Higgses a non-abelian gauge symmetry as
the zero-length strings ending on coincident 5--branes give rise to massless
charged vectors $W_{ij}$.
The 5D gauge coupling $g_5^2$ is inversely proportional to the length
of this stack of coincident branes and becomes infinitely strong for an
infinitesimally short stack \cite{AHK}.
More generally, a loop (or several loops) of branes collapsing to a point in the
56 plane rather than a finite line gives rise to a superconformal field theory in 5D,
similar to a 4--cycle (or several 4--cycles) in M theory collapsing to a single point.

Actually, the five-brane webs in type~IIB superstring theory are dual
to non-compact toric CY3-folds in M theory \cite{LV}.
To see this duality, note that a Calabi--Yau threefold is generally a $T^3$
fibration over $S^3$ or some other real three-fold \cite{MorrisonG}.
In the non-compact case of a resolved singularity such as $\comp^3/\bbbz_N$,
the fibrations base is topologically $\bbbr^3$ or $\bbbr^2\times\bbbr^+$
while the fiber is generically $T^2\times\bbbr$ but
the $T^2$ degenerates to a 1D circle  along a web of real
lines within a single $\bbbr^2$ plane inside the $\bbbr^3$ base.
The duality acts fiber-wise and turns  M theory on a $T^2$ into  type~IIB
superstring on a circle.
A degenerate $T^2$ turns into a IIB five-brane along the locus of degeneration
--- the real line in the base, times the $\bbbr^{4,1}$ Minkowski space ---
whose $(p,q)$ charges depend on the particular 1--cycle $p\alpha_1+q\alpha_2$
of the $T^2$ which shrinks to zero length \cite{LV}.
Thanks to the holomorphy of the Calabi--Yau, the direction of the degeneration
line within the $\bbbr^2$ plane in the base follows the collapsing 1--cycle,
hence the dual $(p,q)$ five-brane satisfies the charge-direction equation~(\ref{Slope})\cite{LV} .

For example, in the $\comp^3/\bbbz_{2M}$ model of the previous section,
the fibration base is shown in the diagram~(\ref{FloorDiagram}):
The $T^2$ degenerates along the solid lines of the diagrams, and the whole
network of these lines is restricted to a single gray surface
(the ``floor''), which topologically is an $\bbbr^2$ plane spanned by
the $|Z_1|^2\pm|Z_2|^2$.
After the $\rm M\leftrightarrow IIB$ duality, the five-brane web looks just
like the floor diagram of~(\ref{FloorDiagram}), namely
\bea
\psset{xunit=8pt,yunit=24pt,linewidth=1pt,arrowscale=1.5}
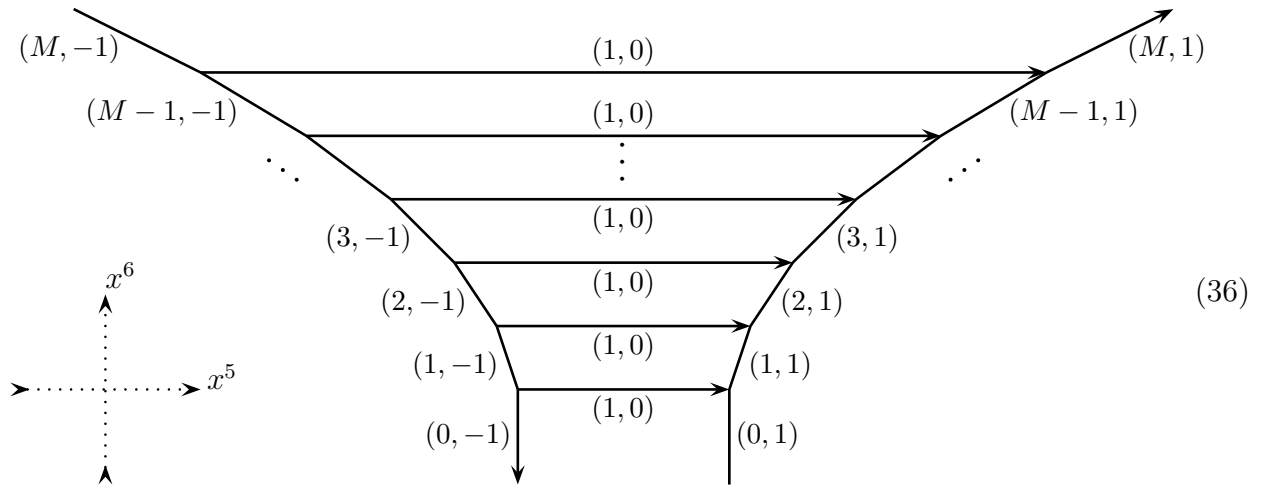
\begin{pspicture}[0.4](-29,-4.5)(+27,+4)
\psline[linestyle=dotted]{>->}(-24.5,-4)(-24.5,-1)
\rput[lb](-24.5,-0.9){$x^6$}
\psline[linestyle=dotted]{>->}(-29,-2.5)(-20,-2.5)
\rput[lb](-19.7,-2.5){$x^5$}
\psline{->}(+5,-4)(+5,-2.5)(+6,-1.5)(+8,-0.5)(+11,+0.5)(+15,+1.5)(+20,+2.5)(+26,+3.5)
\psline{<-}(-5,-4)(-5,-2.5)(-6,-1.5)(-8,-0.5)(-11,+0.5)(-15,+1.5)(-20,+2.5)(-26,+3.5)
\psline{->}(-5,-2.5)(+5,-2.5)
\psline{->}(-6,-1.5)(+6,-1.5)
\psline{->}(-8,-0.5)(+8,-0.5)
\psline{->}(-11,+0.5)(+11,+0.5)
\psline{->}(-15,+1.5)(+15,+1.5)
\psline{->}(-20,+2.5)(+20,+2.5)
\rput[lt](+5.3,-3.0){\small$(0,1)$}
\rput[lt](+5.9,-1.9){\small$(1,1)$}
\rput[lt](+7.4,-0.9){\small$(2,1)$}
\rput[lt](+10.0,+0.1){\small$(3,1)$}
\rput[lt]{35}(+14.8,+0.8){\large${}\cdots{}$}
\rput[lt](+18.2,+2.1){\small\small$(M-1,1)$}
\rput[lt](+23.8,+3.1){\small\small$(M,1)$}
\rput[rt](-5.3,-3.0){\small$(0,-1)$}
\rput[rt](-5.9,-1.9){\small$(1,-1)$}
\rput[rt](-7.4,-0.9){\small$(2,-1)$}
\rput[rt](-10.0,+0.1){\small$(3,-1)$}
\rput[rt]{325}(-14.8,+0.8){\large${}\cdots{}$}
\rput[rt](-18.2,+2.1){\small\small$(M-1,-1)$}
\rput[rt](-23.8,+3.1){\small\small$(M,-1)$}
\rput[b](0,+2.6){\small$(1,0)$}
\rput[b](0,+1.6){\small$(1,0)$}
\rput{90}(0,+1.1){\large${}\cdots{}$}
\rput[t](0,+0.4){\small$(1,0)$}
\rput[t](0,-0.6){\small$(1,0)$}
\rput[t](0,-1.6){\small$(1,0)$}
\rput[t](0,-2.6){\small$(1,0)$}
\end{pspicture}
\label{SUMweb}
\eea
 where 
the $(p,q)$ charges --- and hence the precise directions ---
of the diagonal lines follow from eqs.~(\ref{FiveBraneCharges}).

The {\sl graphic} dual\footnote{%
   The graphic duality maps a planar graph $A$ onto another planar graph
   $B$ such that the loops (faces) of $A$ map onto the vertices of $B$
   and {\it vice verse}, and the corresponding lines of the two graphs
   are perpendicular to each other.
   By abuse of notations, a graphic dual of a five-brane web is the graphic
   dual of its 56 plane.%
   }
of the web~(\ref{SUMweb}) looks like 
\bea
\psset{xunit=24mm,yunit=8mm,linewidth=1pt,linecolor=black}
\begin{pspicture}(-1,-1)(+1,+7)
\psgrid[subgriddiv=1,griddots=10,gridlabels=0](-1,0)(+1,+6)
\psline(-1,0)(0,6)(+1,0)(-1,0)
\psline(-1,0)(0,1)(+1,0)
\psline(-1,0)(0,2)(+1,0)
\psline(-1,0)(0,3)(+1,0)
\psline(-1,0)(0,4)(+1,0)
\psline(-1,0)(0,5)(+1,0)
\psline(0,0)(0,6)
\pscircle[fillstyle=solid,fillcolor=white](0,1){0.12}
\pscircle[fillstyle=solid,fillcolor=white](0,2){0.12}
\pscircle[fillstyle=solid,fillcolor=white](0,3){0.12}
\pscircle[fillstyle=solid,fillcolor=white](0,4){0.12}
\pscircle[fillstyle=solid,fillcolor=white](0,5){0.12}
\pscircle*(0,0){0.1}
\rput[t](-1,-.2){$-1$}
\rput[tl](0,-.25){$0$}
\rput[t](+1,-.2){$+1$}
\rput[rb](-1.2,0){$0$}
\rput[rb](-1.2,1){$1$}
\rput[rb](-1.2,2){$2$}
\rput[rb](-1.2,3){$3$}
\rput[b]{90}(-1.2,4.2){$\cdots$}
\rput[rb](-1.05,5){\small$M-1$}
\rput[rb](-1.15,6){\small$M$}
\end{pspicture}
\label{Triangulated}
\eea
which is clearly the triangulated toric diagram~(\ref{ToricSUM})
of the $\comp^3/\bbbz_{2M}$ orbifold.
This is an example of a general rule:
If type~IIB string theory on a five-brane web~$W$ is dual to M~theory on
a resolved Calabi--Yau singularity $M$, then the graphic dual of the web~$W$ is
the triangulated toric diagram of~$M$.
Indeed, the loops of the web $W$ are dual to the compact 4--cycles of $M$,
which are represented on the toric diagram by internal vertices dual to those loops.
Likewise, the lines separating the loops of $W$ are dual to intersections
of $M$'s 4--cycles, and the graphically dual lines on the toric diagram
connect the corresponding vertices \cite{AHK,LV,KR2}.

\par\vskip 0pt plus 1in \penalty-1000

For example, consider a rank~1 CY singularity of M theory where the collapsing
4--cycle is a Hirzebruch surface $\bbbf_n$.
The toric diagrams of the $\bbbf_n$ are
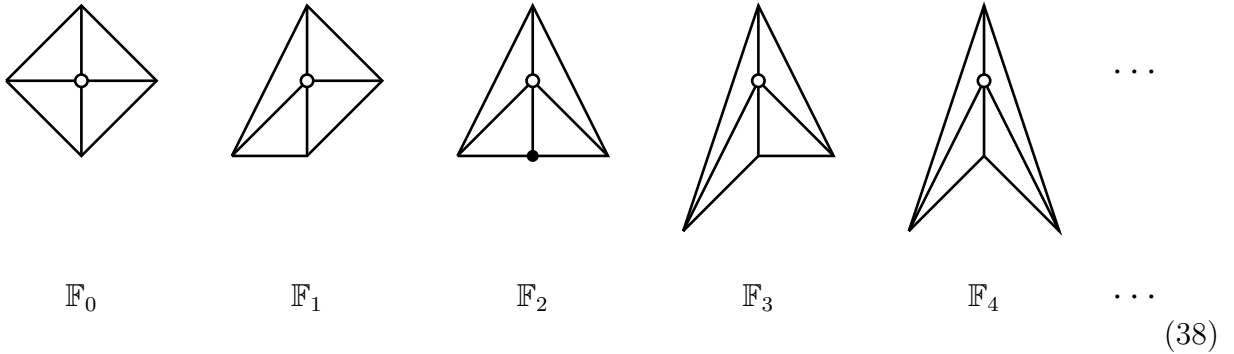
\be
\psset{unit=1cm,linewidth=1pt,linecolor=black}
\begin{pspicture}(-7,-1.5)(+8,+3.5)
\psline(-7,2)(-6,3)(-5,2)(-6,1)(-7,2)
\psline(-7,2)(-5,2)
\psline(-6,1)(-6,3)
\pscircle[fillstyle=solid,fillcolor=white](-6,2){0.1}
\rput[B](-6,-1){$\bbbf_0$}
\psline(-4,1)(-3,3)(-2,2)(-3,1)(-4,1)
\psline(-4,1)(-3,2)(-2,2)
\psline(-3,1)(-3,3)
\pscircle[fillstyle=solid,fillcolor=white](-3,2){0.1}
\rput[B](-3,-1){$\bbbf_1$}
\psline(-1,1)(0,3)(+1,1)(-1,1)
\psline(-1,1)(0,2)(+1,1)
\psline(0,1)(0,3)
\pscircle*(0,1){0.08}
\pscircle[fillstyle=solid,fillcolor=white](0,2){0.1}
\rput[B](0,-1){$\bbbf_2$}
\psline(+2,0)(+3,3)(+4,1)(+3,1)(+2,0)
\psline(+2,0)(+3,2)(+4,1)
\psline(+3,1)(+3,3)
\pscircle[fillstyle=solid,fillcolor=white](+3,2){0.1}
\rput[B](+3,-1){$\bbbf_3$}
\psline(+5,0)(+6,3)(+7,0)(+6,1)(+5,0)
\psline(+5,0)(+6,2)(+7,0)
\psline(+6,1)(+6,3)
\pscircle[fillstyle=solid,fillcolor=white](+6,2){0.1}
\rput[B](+6,-1){$\bbbf_4$}
\rput[B](+8,-1){\large$\cdots$}
\rput[B](+8,+2){\large$\cdots$}
\end{pspicture}
\label{ToricFn}
\ee
and the IIB five-brane webs dual to the resolved CY3-fold singularities form 
graphs dual to the
(\ref{ToricFn}), namely
\be
\psset{unit=4mm,linewidth=1pt,linecolor=black}
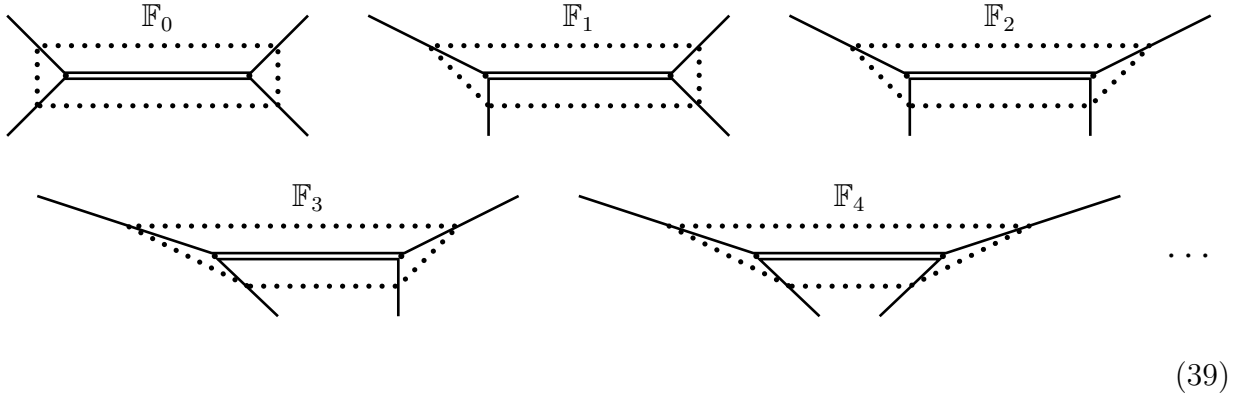
\begin{pspicture}(0,-6.5)(+40,+4.5)
\psline[linestyle=dotted,linewidth=2pt](1,1)(9,1)(9,3)(1,3)(1,1)
\psline(0,0)(1.9,1.9)(8.1,1.9)(10,0)
\psline(0,4)(1.9,2.1)(8.1,2.1)(10,4)
\pscircle*(1.95,2){0.1}
\pscircle*(8.05,2){0.1}
\rput[b](5,3.5){$\bbbf_0$}
\psline[linestyle=dotted,linewidth=2pt](16,1)(23,1)(23,3)(14,3)(16,1)
\psline(16,0)(16,1.9)(22.1,1.9)(24,0)
\psline(12,4)(15.8,2.1)(22.1,2.1)(24,4)
\pscircle*(15.9,2){0.1}
\pscircle*(22.05,2){0.1}
\rput[b](19,3.5){$\bbbf_1$}
\psline[linestyle=dotted,linewidth=2pt](30,1)(36,1)(38,3)(28,3)(30,1)
\psline(30,0)(30,1.9)(36,1.9)(36,0)
\psline(26,4)(29.8,2.1)(36.2,2.1)(40,4)
\pscircle*(29.9,2){0.1}
\pscircle*(36.1,2){0.1}
\rput[b](33,3.5){$\bbbf_2$}
\psline[linestyle=dotted,linewidth=2pt](8,-5)(13,-5)(15,-3)(4,-3)(8,-5)
\psline(9,-6)(7.0,-4.1)(13.0,-4.1)(13,-6)
\psline(1,-2)(6.8,-3.9)(13.2,-3.9)(17,-2)
\pscircle*(6.9,-4){0.1}
\pscircle*(13.1,-4){0.1}
\rput[b](10,-2.5){$\bbbf_3$}
\psline[linestyle=dotted,linewidth=2pt](26,-5)(30,-5)(34,-3)(22,-3)(26,-5)
\psline(27,-6)(25.0,-4.1)(31.0,-4.1)(29,-6)
\psline(19,-2)(24.8,-3.9)(31.2,-3.9)(37,-2)
\pscircle*(24.9,-4){0.1}
\pscircle*(31.1,-4){0.1}
\rput[b](28,-2.5){$\bbbf_4$}
\rput[r](40,-4){\large$\cdots$}
\end{pspicture}
\label{WebFn}
\ee
where the dotted lines describe the Coulomb branch of the 5D theory
and the solid lines the un--Higgsed $SU(2)$ limit.
Note that for $n>2$ the toric diagram is not convex and the web graph
has two converging external lines.
Clearly, these lines cannot extend to infinity, which means the $\bbbf_{n>2}$
web can only be a subset of a bigger web of rank $r>1$, or in the M theory
language, a CY singularity with an $\bbbf_{n>2}$ cycle must have other
compact cycles as well.

For each web diagram~(\ref{WebFn}), the length of the doubled-up
dashed line in the middle of the graph gives the
inverse gauge coupling $h$ of the un-broken 5D $SU(2)$.
We change this coupling by moving the external legs relative to each
other, and for the stand-alone $\bbbf_0$, $\bbbf_1$ and $\bbbf_2$ webs 
the complete parameter space includes the following configurations:
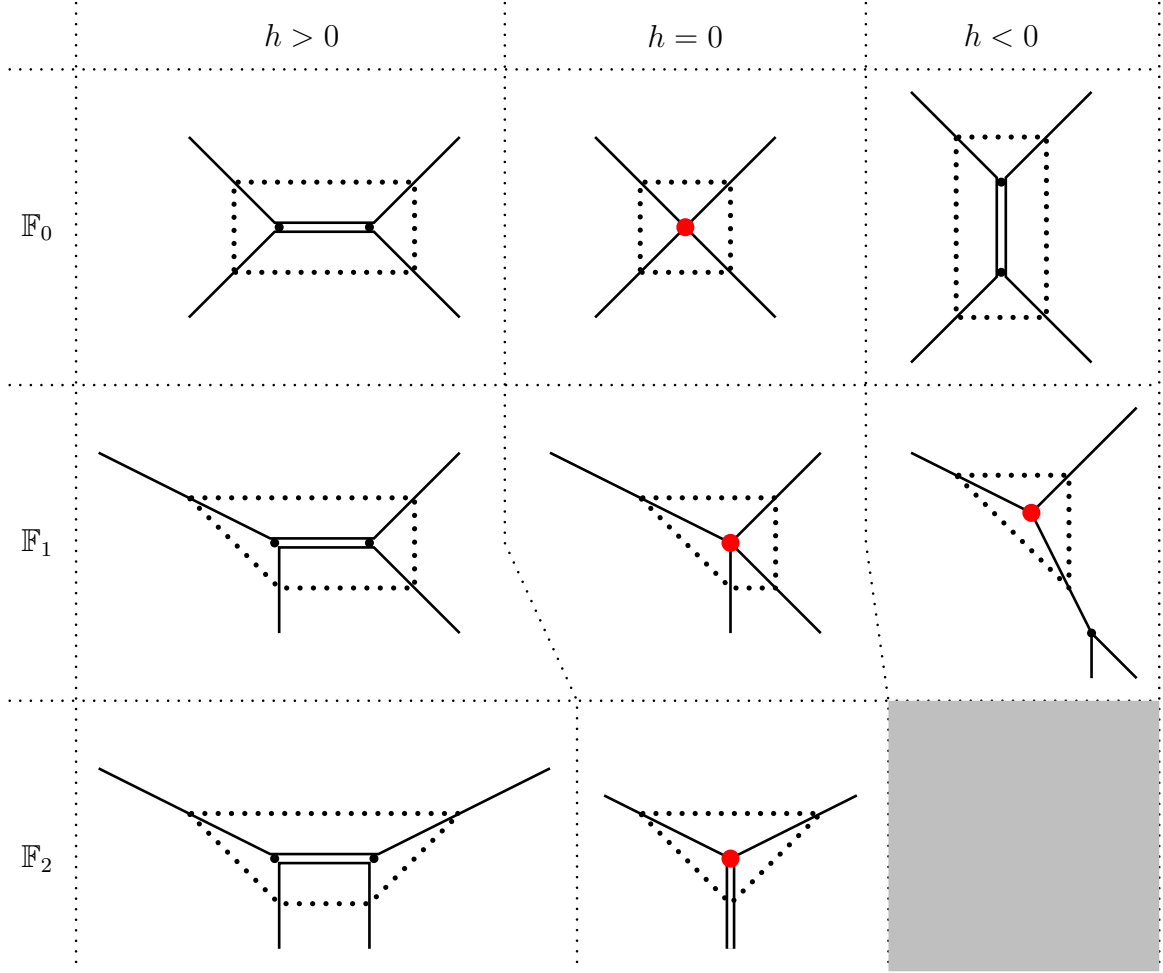
\be
\psset{unit=6mm,linewidth=1pt}
\begin{pspicture}(-3,-10)(23,12.5)
\psline[linestyle=dotted](-1.5,12)(-1.5,-9.5)
\psline[linestyle=dotted](8,12)(8,0)(9.6,-3.5)(9.6,-9.5)
\psline[linestyle=dotted](16,12)(16,0)(16.5,-3.5)(16.5,-9.5)
\psline[linestyle=dotted](22.5,12)(22.5,-9.5)
\rput[b](3.5,11){$h>0$}
\rput[b](12,11){$h=0$}
\rput[b](19,11){$h<0$}
\psline[linestyle=dotted](-3,10.5)(22.5,10.5)
\rput[r](-2,7){$\bbbf_0$}
\psline(1,9)(2.9,7.1)(5.1,7.1)(7,9)
\psline(1,5)(2.9,6.9)(5.1,6.9)(7,5)
\pscircle*(3,7){0.1}
\pscircle*(5,7){0.1}
\psline[linestyle=dotted,linewidth=2pt](2,6)(2,8)(6,8)(6,6)(2,6)
\psline(10,9)(12,7)(14,9)
\psline(10,5)(12,7)(14,5)
\pscircle*[linecolor=red](12,7){0.2}
\psline[linestyle=dotted,linewidth=2pt](11,6)(11,8)(13,8)(13,6)(11,6)
\psline(17,10)(18.9,8.1)(18.9,5.9)(17,4)
\psline(21,10)(19.1,8.1)(19.1,5.9)(21,4)
\pscircle*(19,8){0.1}
\pscircle*(19,6){0.1}
\psline[linestyle=dotted,linewidth=2pt](18,9)(20,9)(20,5)(18,5)(18,9)
\psline[linestyle=dotted](-3,3.5)(22.5,3.5)
\rput[r](-2,0){$\bbbf_1$}
\psline(-1,+2)(2.8,+0.1)(5.1,+0.1)(7,+2)
\psline(3,-2)(3,-0.1)(5.1,-0.1)(7,-2)
\pscircle*(2.9,0){0.1}
\pscircle*(5,0){0.1}
\psline[linestyle=dotted,linewidth=2pt](1,+1)(6,+1)(6,-1)(3,-1)(1,+1)
\psline(9,+2)(13,0)(15,+2)
\psline(13,-2)(13,0)(15,-2)
\pscircle*[linecolor=red](13,0){0.2}
\psline[linestyle=dotted,linewidth=2pt](11,+1)(14,+1)(14,-1)(13,-1)(11,+1)
\psline(17,+2)(19.667,+0.667)(22,+3)
\psline(19.667,+0.667)(21,-2)(21,-3)
\psline(21,-2)(22,-3)
\pscircle*(21,-2){0.1}
\pscircle*[linecolor=red](19.667,+0.667){0.2}
\psline[linestyle=dotted,linewidth=2pt](18,+1.5)(20.5,+1.5)(20.5,-1)(18,+1.5)
\psline[linestyle=dotted](-3,-3.5)(22.5,-3.5)
\rput[r](-2,-7){$\bbbf_2$}
\psline(-1,-5)(2.8,-6.9)(5.2,-6.9)(9,-5)
\psline(3,-9)(3,-7.1)(5,-7.1)(5,-9)
\pscircle*(2.9,-7){0.1}
\pscircle*(5.1,-7){0.1}
\psline[linestyle=dotted,linewidth=2pt](1,-6)(7,-6)(5,-8)(3,-8)(1,-6)
\psline(10.2,-5.6)(13,-7)(15.8,-5.6)
\psline[doubleline=true,doublesep=0.1](13,-7)(13,-9)
\pscircle*[linecolor=red](13,-7){0.2}
\psline[linestyle=dotted,linewidth=2pt](11,-6)(15,-6)(13,-8)(11,-6)
\psframe*[linecolor=lightgray](16.5,-9.5)(22.5,-3.5)
\end{pspicture}
\label{SUtwoFlops}
\ee
where the red circles
\enspace\raise 0.5ex \hbox{\pscircle*[linecolor=red]{1.2mm}}\enspace\space
 denote loops collapsing to points
and giving rise to superconformal theories in 5D.
We see that all three webs have a superconformal point at $h=0$, but the
flop transition to $h<0$ is very different:
For the $\bbbf_0$, the flop is a symmetry and
the same $SU(2)$ SYM with $\theta=0$ obtains for both $h>0$ and $h<0$.
The $\bbbf_1$ has two different phases: The $SU(2)$ SYM with $\theta=\pi$
for $h>0$ and the $E_0$ superconformal theory (the red circle)
plus some massive particles for $h<0$.
Finally, the $\bbbf_2$ web does not flop at all: $h$ is the distance between two
parallel external legs, and it cannot go negative no matter what.
This confirms the M--theory result: The parameter space of the $\bbbf_2$ singularity
is $\bbbr^+$ rather than $\bbbr$ and there are no flop transitions.

Now consider the $SU(M)$ theories with $M>2$.
The five-brane web (\ref{SUMweb}) dual to the $\comp^3/\bbbz_{2M}$ orbifold of M theory
behaves exactly line the $\bbbf_2$ web for $M=2$:
A quick look at the web of the un--Higgsed $SU(M)$
\be
\psset{xunit=2.8mm,yunit=8.4mm,linewidth=1pt}
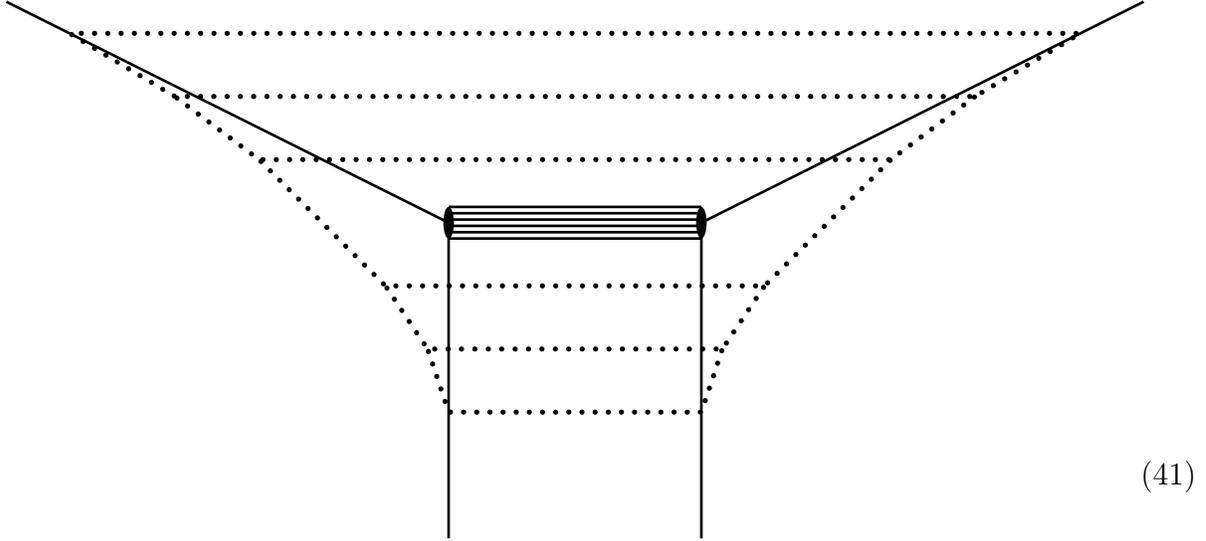
\begin{pspicture}[0.1](-24,-5)(+24,+3.5)
\psline(-27,+3.5)(-6,0)(-6,-5)
\psline(+27,+3.5)(+6,0)(+6,-5)
\psline(-6,-0.25)(+6,-0.25)
\psline(-6,-0.15)(+6,-0.15)
\psline(-6,-0.05)(+6,-0.05)
\psline(-6,+0.05)(+6,+0.05)
\psline(-6,+0.15)(+6,+0.15)
\psline(-6,+0.25)(+6,+0.25)
\psellipse*(-6,0)(0.25,0.25)
\psellipse*(+6,0)(0.25,0.25)
\psline[linestyle=dotted,linewidth=2pt](-7,-2)(-6,-3)(+6,-3)(+7,-2)
\psline[linestyle=dotted,linewidth=2pt](-9,-1)(-7,-2)(+7,-2)(+9,-1)
\psline[linestyle=dotted,linewidth=2pt](-15,+1)(-9,-1)(+9,-1)(+15,+1)
\psline[linestyle=dotted,linewidth=2pt](-19,+2)(-15,+1)(+15,+1)(+19,+2)
\psline[linestyle=dotted,linewidth=2pt](-24,+3)(-19,+2)(+19,+2)(+24,+3)(-24,+3)
\end{pspicture}
\label{SUMunHiggsed}
\ee
shows that the inverse gauge coupling $h$ is the distance between two
parallel external legs, hence $h\ge0$ no matter what and $\bbbr^+$
parameter space.
But just as in the $M=2$ case, there are other webs which allow flop
transitions between an $SU(M)$ SYM for $h>0$ and some other 5D theory
for $h<0$.
Generally, for $h>0$ all such webs are $SL(2,\bbbz)$ equivalent to
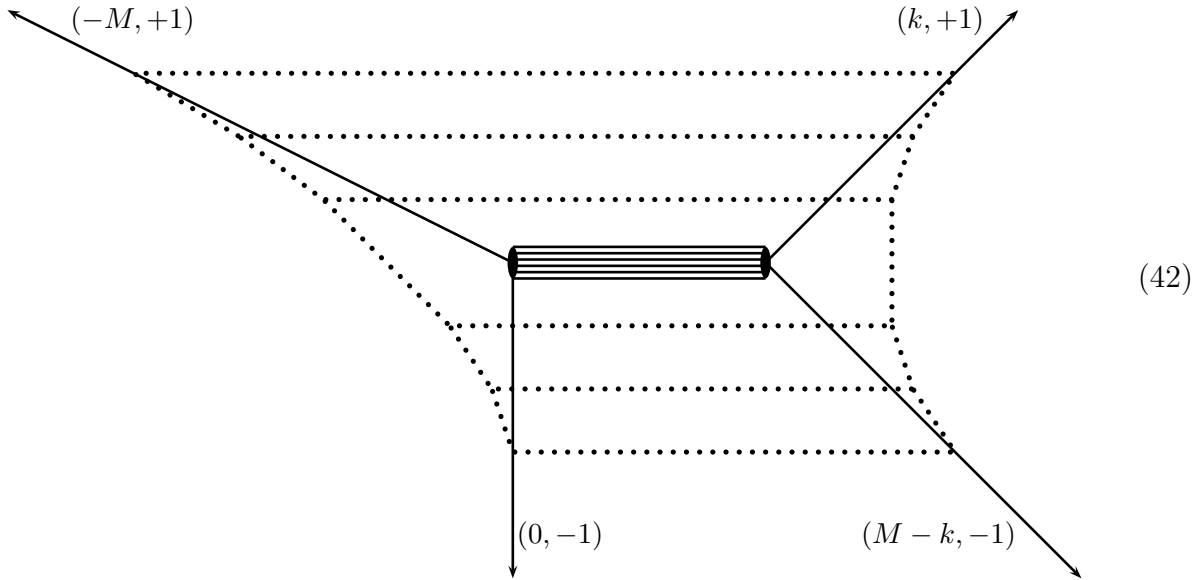
\be
\psset{xunit=2.8mm,yunit=8.4mm,linewidth=1pt}
\begin{pspicture}[0.5](-30,-5.25)(+21,+4.5)
\psline{<-}(-30,+4)(-6,0)
\psline{<-}(+18,+4)(+6,0)
\psline{->}(-6,0)(-6,-5)
\psline{->}(+6,0)(+21,-5)
\rput[lb](-27,+3.6){\small$(-M,+1)$}
\rput[rb](+16.4,+3.6){\small$(k,+1)$}
\rput[lt](-5.8,-4.1){\small$(0,-1)$}
\rput[rt](+17.9,-4.1){\small$(M-k,-1)$}
\psline(-6,-0.25)(+6,-0.25)
\psline(-6,-0.15)(+6,-0.15)
\psline(-6,-0.05)(+6,-0.05)
\psline(-6,+0.05)(+6,+0.05)
\psline(-6,+0.15)(+6,+0.15)
\psline(-6,+0.25)(+6,+0.25)
\psellipse*(-6,0)(0.25,0.25)
\psellipse*(+6,0)(0.25,0.25)
\psline[linestyle=dotted,linewidth=2pt](-7,-2)(-6,-3)(+15,-3)(+13,-2)
\psline[linestyle=dotted,linewidth=2pt](-9,-1)(-7,-2)(+13,-2)(+12,-1)
\psline[linestyle=dotted,linewidth=2pt](-15,+1)(-9,-1)(+12,-1)(+12,+1)
\psline[linestyle=dotted,linewidth=2pt](-19,+2)(-15,+1)(+12,+1)(+13,+2)
\psline[linestyle=dotted,linewidth=2pt](-24,+3)(-19,+2)(+13,+2)(+15,+3)(-24,+3)
\end{pspicture}
\label{SUMwebKCS}
\ee
for some integer $k$ between $-M$ and $+M$,
and two parallel external legs --- and hence no flop transitions to $h<0$ ---
only for $k=\pm M$.

Physically, $k$ is the Chern--Simons level of the 5D theory.
Indeed, consider the Coulomb branch of the web depicted by the dotted lines
on fig.~(\ref{SUMwebKCS}).
Let $y_1<y_2<\cdots y_M$ denote the $x^6$ positions of the $(1,0)$ branes
(the vertical positions of the horizontal lines)
and $t_1,t_2,\ldots,t_M$ their lengths in the $x^5$ direction;
thanks to eqs.~(\ref{FiveBraneCharges}),
\be
t_i\ =\ h\ +\ k\phi_i\ +\sum_{j=1}^M\left|y_i-y_j\right| .
\label{SUMwebCouplings}
\ee
From the 5D point of view, the $y_i$ are the eigenvalues $\phi_i$ of the adjoint
scalar field breaking the $SU(M)\to U(1)^{M-1}$ while the $t_i$
are the inverse gauge couplings of the surviving abelian symmetries.
Thus, we interpret the web geometry relations~(\ref{SUMwebCouplings})
in terms of the 5D gauge coupling formulae~(\ref{SUMGaugeCouplings}),
which immediately identifies $k$ as the Chern--Simons level~$\kcs$.

{\bf For $k\neq\pm M$}, the external legs of the web diagram~(\ref{SUMwebKCS})
are diverging rather than parallel and {\bf there is a flop transition to $h<0$.}
As an example, consider the case of $SU(3)$ at $\kcs=2$ where the web
flops to an exotic rank~2 SCFT (plus massive junk):
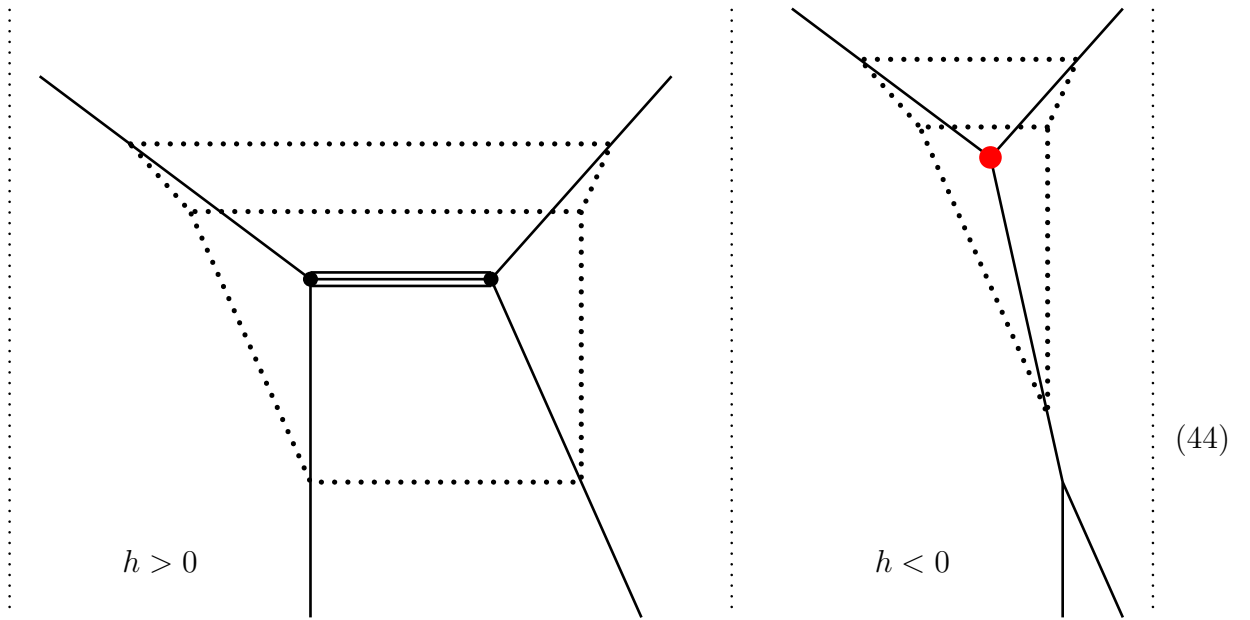
\be
\psset{xunit=4mm,yunit=9mm,linewidth=1pt}
\begin{pspicture}[0.3](-13,-5.5)(+25,+4.5)
\psline(-12,+3)(-3,0)(-3,-5)
\psline(+9,+3)(+3,0)(+8,-5)
\psline(-3,+0.1)(+3,+0.1)
\psline(-3,0)(+3,0)
\psline(-3,-0.1)(+3,-0.1)
\pscircle*(-3,0){0.1}
\pscircle*(+3,0){0.1}
\psline[linestyle=dotted,linewidth=2pt](-9,+2)(-7,+1)(+6,+1)(+7,+2)(-9,+2)
\psline[linestyle=dotted,linewidth=2pt](-7,+1)(-3,-3)(+6,-3)(+6,+1)
\psline[linestyle=dotted](11,+4)(11,-5)
\psline[linestyle=dotted](-13,+4)(-13,-5)
\psline[linestyle=dotted](25,+4)(25,-5)
\rput[t](-8,-4){$h>0$}
\rput[t](17,-4){$h<0$}
\psline(13,+4)(19.6,+1.8)(22,-3)(22,-5)
\psline(19.6,+1.8)(24,+4)
\psline(22,-3)(24,-5)
\pscircle*[linecolor=red](19.6,+1.8){0.15}
\psline[linestyle=dotted,linewidth=2pt]%
    (17.25,+2.25)(15.25,+3.25)(22.5,+3.25)(21.5,+2.25)(21.5,-2)(17.25,+2.25)(21.5,+2.25)
\end{pspicture}
\label{SUthreeFlop}
\ee
In the dual M theory, this flop changes the triangulation of the toric diagram
of the CY singularity:
\be
\psset{unit=12mm,linewidth=1pt,arrowscale=2}
\begin{pspicture}[0.125](0,-0.5)(7,3.5)
\psline(0,0)(1,3)(2,1)(1,0)(0,0)
\psline(0,0)(1,2)(2,1)(1,1)(0,0)
\psline(1,3)(1,0)
\pscircle[fillcolor=white,fillstyle=solid](1,2){0.1}
\pscircle[fillcolor=white,fillstyle=solid](1,1){0.1}
\psline(5,0)(6,3)(7,1)(6,0)(5,0)
\psline(5,0)(6,2)(7,1)(6,1)(5,0)
\psline(6,3)(6,1)
\psline(5,0)(7,1)
\pscircle[fillcolor=white,fillstyle=solid](6,2){0.1}
\pscircle[fillcolor=white,fillstyle=solid](6,1){0.1}
\psline[linewidth=0.6pt]{<->}(2.5,1.5)(4.75,1.5)
\rput[tr](0,3){$h>0$}
\rput[tl](7,3){$h<0$}
\end{pspicture}
\label{SUthreeToric}
\ee
Focusing on the 5D SCFT for $h<0$ and ignoring the massive stuff,
we truncate the toric diagram and the web to
\be
\psset{unit=12mm,linewidth=1pt}
\begin{pspicture}[0.5](0,0)(3,3)
\psline(0,0)(1,3)(2,1)(0,0)(1,2)(2,1)(1,1)(0,0)
\psline(1,3)(1,1)
\pscircle[fillcolor=white,fillstyle=solid](1,2){0.1}
\pscircle[fillcolor=white,fillstyle=solid](1,1){0.1}
\end{pspicture}
\psset{unit=4.5mm}
\begin{pspicture}[0.5](-1,-4.5)(12,4.5)
\psline(0,4)(3,3)(5,2)(8,-2)(9,-4)
\psline(11,4)(9,3)(8,2)(8,-2)
\psline(3,3)(9,3)
\psline(5,2)(8,2)
\end{pspicture}
\label{Zfive}
\ee
In M theory, this describes 2 compact 4--cycles --- an $\bbbf_3$
and a $\bbbp^2$ --- intersecting each other along a $\bbbp^1$.
The triangular shape of the toric diagram indicates that
the singularity containing these cycles is an orbifold fixed point;
an $SL(2,\bbbz)$ transform maps the triangle's vertices to respectively
$(0,1)$, $(1,0)$ and $(-1,-3)$, which pinpoint a particular orbifold,
namely the $\comp^3/\bbbz_5[1,1,3]$.
On the IIB side, the truncated web has only three external lines and thus
no non-dynamical parameters (the $h<0$ parameter of the full web affects
the massive junk but decouples from the SCFT in question).
Hence, when both loops of the web graph shrink in area, they collapse
to a single point rather than a stack of coincident lines.
In M theory terms, this means the singularity cannot be partially
resolved into a line; it has to be resolved in full or not at all.
And in 5D terms, this means an exotic SCFT which does not follow from 
the $g_5=\infty$ limit of an ordinary gauge theory.

Note that the $\comp^3/\bbbz_5$ orbifold in M theory describes the
SCFT degrees of freedom of the $h<0$ phase of the $SU(3)$
(at $\kcs=2$) but without additional particles with $h$-dependent
masses one cannot flop back into the $h>0$ SYM phase.
The full 5D theory follows from M theory on a more complicated
singularity whose toric diagram is (\ref{SUthreeToric})
rather than (\ref{Zfive}), and this singularity is not of the
orbifold type.
Indeed, for all of the $SU(M)$ theories at Chern--Simons levels $k\neq\pm M$,
embedding into M theory requires a more complicated CY singularity than just
a $\comp^3/\bbbz_N$ orbifold because the
the toric diagrams  dual to their (\ref{SUMwebKCS}) webs are proper quadrangles
rather than triangles.
The $\kcs=\pm M$ theories are fortunate exceptions from this rule.

Conversely, M theory on
orbifold singularities with eigenvalues other than $(1,1,2M-2)$
produce 5D theories more complicated than a non-abelian SYM.
For example, for odd $N=2M+1$, the $\comp^3/\bbbz_N$ orbifold
with eigenvalues $(1,1,2M-1)$ is an exotic SCFT of rank $M$:
The good old $E_0$ for the $\bbbz_3$ orbifold, the rank~2
theory depicted on fig.~(\ref{Zfive}) for the $\bbbz_5$, or
a higher-rank generalization of the same pattern for odd $N>5$:
The compact 4--cycles are $\{\bbbf_{N-2},\bbbf_{N-4},\cdots,\bbbf_3,\bbbp^2\}$,
and when all the cycles shrink, they collapse to a single point rather
than a line.
Another example is the $\bbbz_{12}$ orbifold with eigenvalues $(1,4,7)$ whose
toric diagram is shown in fig.~(\ref{ToricExample});
its type~IIB brane web dual follows via graphic duality:
Modulo a choice of triangulations and an $SL(2,\bbbz)$ rotation,
the web looks like
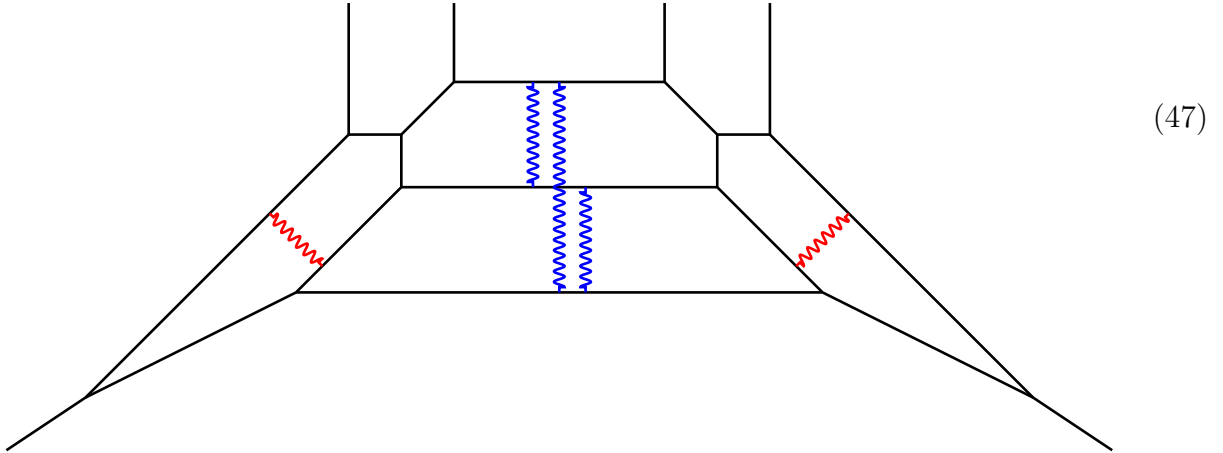
\be
\psset{unit=7mm,linewidth=1pt,coilwidth=0.2,coilaspect=0}
\begin{pspicture}[0.7](-10.5,-3.5)(+10.5,+6)
\psline(-10.5,-3)(-9,-2)(-4,+3)(-4,+5.5)
\psline(-9,-2)(-5,0)(-3,+2)(-3,+3)(-2,+4)(-2,+5.5)
\psline(-4,+3)(-3,+3)
\psline(+10.5,-3)(+9,-2)(+4,+3)(+4,+5.5)
\psline(+9,-2)(+5,0)(+3,+2)(+3,+3)(+2,+4)(+2,+5.5)
\psline(+4,+3)(+3,+3)
\psline(-2,+4)(+2,+4)
\psline(-3,+2)(+3,+2)
\psline(-5,0)(+5,0)
\pscoil[coilarm=0.1,linecolor=blue](-0.5,+4)(-0.5,+2)
\pscoil[coilarm=0.1,linecolor=blue](+0.5,+2)(+0.5,0)
\pscoil[coilarm=0.1,linecolor=blue](0,0)(0,+4)
\pscoil[coilarm=0.1,linecolor=red](-4.5,+0.5)(-5.5,+1.5)
\pscoil[coilarm=0.1,linecolor=red](+4.5,+0.5)(+5.5,+1.5)
\end{pspicture}
\ee
where the colored wavy lines indicate strings between
parallel five-branes which generate the ${\blue SU(3)}\times
{\red SU(2)}\times{\red SU(2)}$ non-abelian gauge symmetry.
Altogether, the model has 3 non-dynamical coupling/mass parameters and
4 dynamical moduli, but their roles are mixed up:
One of the two moduli controlling the $\blue SU(3)$ breaking also controls
the gauge couplings of the two $\red SU(2)$ factors.
Consequently, one cannot un--Higgs the entire ${\blue SU(3)}\times
{\red SU(2)}\times{\red SU(2)}$ symmetry
while keeping all the gauge couplings finite.

We have other examples of weird 5D theories following from M theory
orbifolds $\comp^3/\bbbz_N$, and we hope to present the whole menagerie
in a future publication.
For the moment, we return to the main subject of this article --- the $SU(M)$
SYM theories in 5D and their  moduli/parameter spaces.
We have learned how everything works in the M/string based UV completions
of the 5D SYM, so let us consider the alternative UV completion via
dimensional deconstruction.

\section{Deconstruction}
In this section, we use dimensional deconstruction to study the quantum
moduli and parameter spaces of the 5D $SU(M)$ SYM theory.
That is, we deconstruct the fifth dimension of the theory and replace
it with a discrete lattice of small bit finite spacing $a$.
This breaks the Lorentz symmetry down to $\rm Spin(3,1)$ and also
breaks 4 out of 8 supercharges of the theory.
The result is a 4D, ${\cal N}=1$ gauge theory on the following quiver:
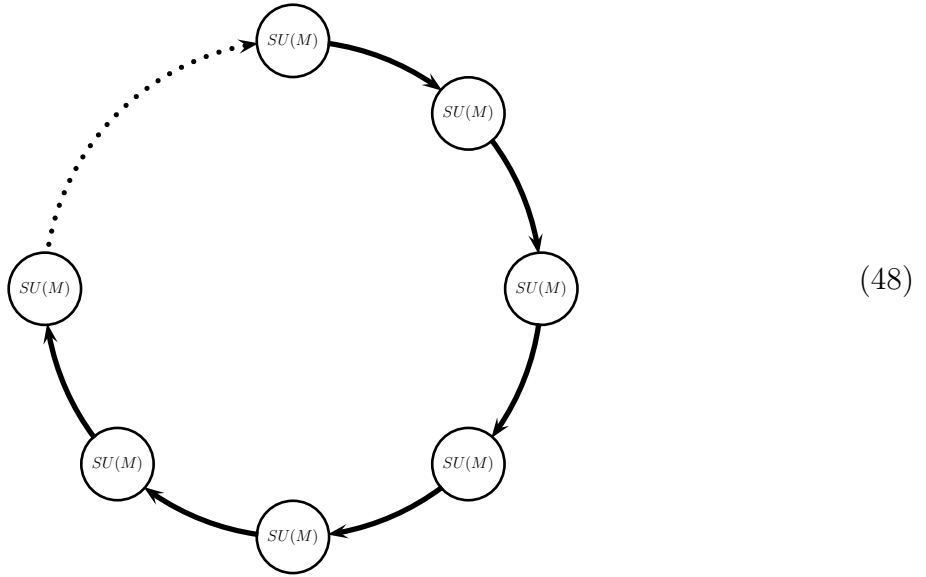
\be
\def\SUM{\scalebox{0.5}{$SU(M)$}}
\psset{unit=33mm,linewidth=1pt,linecolor=black}
\begin{pspicture}[0.5](-1.2,-1.2)(+1.2,+1.2)
\cput(0,+1){\SUM}
\cput(+.707,+.707){\SUM}
\cput(+1,0){\SUM}
\cput(+.707,-.707){\SUM}
\cput(0,-1){\SUM}
\cput(-.707,-.707){\SUM}
\cput(-1,0){\SUM}
\psarc[linewidth=2pt]{<-}(0,0){1}{53}{82}
\psarc[linewidth=2pt]{<-}(0,0){1}{8}{37}
\psarc[linewidth=2pt]{<-}(0,0){1}{323}{352}
\psarc[linewidth=2pt]{<-}(0,0){1}{278}{307}
\psarc[linewidth=2pt]{<-}(0,0){1}{233}{262}
\psarc[linewidth=2pt]{<-}(0,0){1}{188}{217}
\psarc[linewidth=2pt,linestyle=dotted]{<-}(0,0){1}{98}{172}
\end{pspicture}
\label{QuiverDiagram}
\ee
That is, the 4D gauge symmetry is
\be
G_4\ =\ \prod_{\ell=1}^N SU(M)_\ell
\label{QuiverG}
\ee
while the chiral superfields comprise $N$ bi-fundamental multiplets
\be
Q_\ell\ =\ ({\bf M}_\ell, \overline{\bf M}_{\ell+1}),\,\,\,\,
\ell=1,2,\ldots,N\ \mbox{modulo}\, N.
\label{QuiverQ}
\ee
Classically, the scalars in $Q_\ell$ are related to the 5D fields
according to
\be
Q_\ell\
=\ V\times\left[ {\bf1\!\!\!1}\,
+\, a\times\left(\phi+iA^5\right)_{x=\ell a}\,
+\, O(a^2\phi^2)\right]
\label{QuiverVEV}
\ee
for some fixed $V\neq0$.

In a vacuum state ($\phi+iA^5=\rm const$ modulo gauge equivalence,
$[\phi,A^5]=0$), the gauge symmetry (\ref{QuiverG}) is spontaneously
broken down to the diagonal $SU(M)$ (for $\phi+iA^5=0$) or a subgroup
thereof; for generic $\phi+iA^5$, only the diagonal Cartan subgroup
$U(1)^{M-1}\subset SU(M)_{\rm diag}$ survives unbroken.
All the remaining vector fields acquire tree-level masses
\be
M^2(W_k^{ij})\ 
=\ g^2|V|^2(1+O(a\phi))\times\left[
     4\sin^2\frac{1}{2}\left(\frac{2\pi k}{N}+a(A^5_i-A^5_j)\right)\,
    +\,(\phi_i-\phi_j)^2 a^2(1+O(a\phi))
    \right]
\label{QuiverMasses}
\ee
where $i$ and $j$ are $SU(M)$ indices (in the eigenbasis of $\phi$ and $A^5$)
and $k$ is discrete Fourier transform of the quiver index~$\ell$.
Let us identify \cite{csaki,csaki2}
\be
g^2 |V|^2\ =\ \frac{1}{a^2} ;
\ee
then for weak fields $\phi\ll (1/a)$ and low 5-momenta $k\ll N$,
the spectrum (\ref{QuiverMasses}) becomes exactly the Kaluza--Klein
spectrum
\be
M^2\ =\ (\phi_i-\phi_j)^2\ +\ p_5^2 ,\qquad
p_5^{}\ =\ \frac{2\pi k}{Na}\,+\,(A^5_i-A^5_j)
\label{KaluzaKlein}
\ee
of the Coulomb branch of the 5D theory compactified on a circle
of length $L=Na$.
In the large quiver limit $N\to\infty$, the fifth dimension un-compactifies
and the spectrum becomes continuous.

\subsection{Deconstructing the Gauge Couplings}
In the quantum theory of the quiver~(\ref{QuiverDiagram}),
the masses~(\ref{QuiverMasses}) are corrected by all kinds of
perturbative and non-perturbative effects, and we do not have
any tools for deriving an exact formula.
Instead, we focus on the holomorphic gauge couplings $\tau_{ij}$ of
the unbroken 4D gauge symmetry $U(1)^{M-1}$.
We shall calculate these couplings exactly and show that in the
large quiver limit $N\to\infty$ they behave as
\be
2\pi\Im \tau_{ij}\,\equiv\,\left[\frac{8\pi^2}{g_4^2}\right]_{ij}\
=\ Na\times \left[\frac{8\pi^2}{g_5^2}\right]_{ij}\ +\ O(1),
\label{GFourFive}
\ee
in complete agreement with the Kaluza--Klein reduction of a 5D
gauge theory.
Furthermore, the 5D gauge coupling matrix on the right hand side
of eq.~(\ref{GFourFive}) are exactly as in
eq.~(\ref{SUMGaugeCouplings}) for the un-deconstructed 5D $SU(M)$ SYM.

Our first step is to relate the 5D eigenvalues $\phi_i$ to the
holomorphic gauge-invariant moduli of the quiver theory.
Altogether, the Coulomb branch of the quiver has
\be
\#(\mbox{Chiral SF})\ -\ \#(\mbox{Broken Symmetries})\
=\ N+M-1
\ee
independent moduli, for example the ``baryon'' fields $B_\ell=\det(Q_\ell)$,
and the coefficients $T_1,T_2,\ldots,T_{M-1}$ of the characteristic polynomial
\bea
\label{CharPol}
T(X) &\equiv& \Det\left(X-Q_{1}Q_{2}\cdots Q_{N}\right) \\
\nonumber
&=&\ X^M-T_{1}X^{M-1}+T_{2}X^{M-2}-\cdots\mp T_{M-1}X\pm T_M
\eea
of the quiver-ordered matrix product of the bilinear fields $Q_\ell$.
The classical relation~(\ref{QuiverVEV}) between the $Q_\ell$ and
the 5D fields yields $B_\ell=V^M$ and
\bea
T(X)\ &=&\ \prod_{i=1}^M\left(X\,-\,V^N\exp(Na\varphi_i)\right) \\
\mbox{where}\ \varphi_i\ &=&\ \phi_i\,+\,iA^5_i\,+\,O(a\phi^2) .
\nonumber
\eea
Consequently, in the quantum quiver theory, we {\sl define}
\be
\varphi_i\,\equiv\,\phi_i+iA^5_i\
\equiv\ \frac{1}{Na}\log\frac{R_i}{V^N}
\label{PhiDef}
\ee
where $R_i$ are the roots of the characteristic polynomial
$T(X)=\prod_i(X-R_i)$.
In order to assure $\sum_i\varphi_i=0$ as required by the $SU(M)$ theory,
we {\sl define}  $V$ according to
\be
V^{MN}\ =\ \prod_{i=1}^M R_i\ =\ T_M .
\label{Vdef}
\ee

Note that $T_M$ is not an independent modulus of the quiver
but a function of the baryon moduli $B_\ell$ \cite{seibergq,paper1}.
Classically,
\be
T_M\equiv\left\langle\Det(Q_1Q_2\cdots Q_N)\right\rangle
=\prod_{\ell=1}^N\left(\left\langle Q_\ell\right\rangle\equiv B_\ell\right) ,
\ee
but quantum corrections \cite{rodriguez,georgi} modify this
relation to
\bea
T_M &=& \mathop{\rm Polynomial\ part\ of}\left[
        \prod_\ell B_\ell\times
        \prod_\ell\left(1+{\Lambda^{2M}_\ell\over B_\ell B_{\ell+1}}\right)
        \right] \hbox{\vrule width 0pt depth 20pt}\\
\nonumber &=&
\prod_\ell B_\ell \times \left\{
    \vcenter{\openup 1\jot\ialign{%
        \hfil ${}#$\ & $\displaystyle{{}+#\vphantom{\sum}}$\hfil \cr
        1& \sum_\ell\frac{\Lambda^{2M}_\ell}{ B_\ell B_{\ell+1}}\,
            +\,\frac{1}{2}\sum_\ell\;\sum_{\ell'\neq \ell,\ell\pm1}
                \frac{\Lambda^{2M}_\ell}{B_\ell B_{\ell+1}}\,
                        \frac{\Lambda^{2M}_{\ell'}}{ B_{\ell'} B_{k\ell'+1}}\cr
        &\,\frac{1}{6}\sum_\ell\;\sum_{\ell'\neq \ell,\ell'\pm1}\;\;
                                        \sum_{\ell''\neq \ell,\ell\pm1,\ell',\ell'\pm1}
            \frac{\Lambda^{2M}_\ell}{B_\ell B_{\ell+1}}\,
                        \frac{\Lambda^{2M}_{\ell'}}{B_{\ell'} B_{\ell'+1}}\,
                                \frac{\Lambda^{2M}_{\ell''}}{B_{\ell''} B_{\ell''+1}}\cr
        &\ \cdots\cr
    }}\right\} \\
\label{QuantumCorr}
\eea
For the deconstruction purposes we are interested in translationally
($\bbbz_N$) invariant quivers with equal $\Lambda_\ell\equiv\Lambda$.
Likewise, we fix the baryon moduli to equal values $B_\ell\equiv B$.%
\footnote{%
        The simplest way to fix all the baryonic moduli is to
        endow the quiver with a tree-level superpotential of the form
        \be
        W=\sum_\ell\left(\det(Q_\ell)-B\right)\times S_\ell
        \ee
        where $S_\ell$ are some additional gauge-singlet chiral superfields.
        Although this superpotential is non-renormalizable for $M>2$,
        the resulting divergences of the 4D theory can be regularized
        using covariant higher-derivative Lagrangian terms without affecting
        any of the holomorphic properties of the quiver.
        \par
        Alternatively, we may leave the baryons un-fixed and gauge
        the $U(1)^N$ flavor symmetry of the quiver.
        {}From the 5D point of view, this would mean converting a
        non-dynamical parameter $h$ of
        the prepotential~(\ref{SUMPrepotential})
        into a dynamical vector multiplet.
        }
Consequently, eq.~(\ref{QuantumCorr}) yields
\be
T_M\
= \sum_{n=0}^{\mathop{\rm int}[N/2]}\,
\frac{N\,(N-n-1)!}{n!\,(N-2n)!}\,B^{N-2n}\Lambda^{2Mn}\,,
\ee
which sums to
\be
T_M\ =\ v_1^{MN}\ +\ v_2^{MN}
\label{TMformula}
\ee
where $v_{1,2}^M=\frac{1}{2}(B\pm\sqrt{B^2+4\Lambda^{2M}})$
are the two roots of the quadratic
equation system
\be
v_1^M\,+\,v_2^M\ =\ B,\qquad
v_1^M\,\times\,v_2^M\ =\ \Lambda^{2M} .
\label{Vroots}
\ee
In the large quiver limit $N\to\infty$, the right hand side of eq.~(\ref{TMformula})
is completely dominated by the larger of the two terms there,
hence
\be
\lim_{N\to\infty} T_M\ =\ \max(v_1^{MN},v_2^{MN}) ,
\ee
and in light of eq.~(\ref{Vdef}),
\be
V\ =\ \max(v_1,v_2) .
\label{Vformula}
\ee
Note that according to eqs.~(\ref{Vroots}) and (\ref{Vformula}),
\be
|V|\ \ge\ |\Lambda|
\label{Vbound}
\ee
regardless of the baryon VEV~$B$.
Although $V$ is a holomorphic parameter of the quiver, in the large $N$ limit
there is a discontinuity which keeps $V$ outside a complex circle of radius $|\Lambda|$,
and any attempt to analytically continue $V$ inside that circle results in a
root exchange $v_1\leftrightarrow v_2$ and hence inversion
$V\to \Lambda^2/V$ back to the outside of the $|\Lambda|$ circle.

Now that we understand the moduli of the quiver theory, let us consider
the moduli dependence of the gauge couplings of its Coulomb branch.
Logically, the gauge symmetry of the quiver is broken in two stages,
\be
\left[SU(M)\right]^N\ \to\ SU(M)_{\rm diag}\
\to\ U(1)^{M-1} ,
\ee
where the second stage breaking is due to an effective adjoint multiplet
of the diagonal $SU(M)$.
Consequently, the quiver has a Seiberg--Witten hyperelliptic curve similar
to that of an $SU(M)$ theory with an adjoint matter multiplet $A$,
namely
\be
Y^2\ =\ \left[\Det(X-A)\right]^2\ -\ 4\Lambda^{2M} .
\ee
For the quiver, the the role of the adjoint $A$ is played by the
quiver-ordered product $Q_1Q_2\cdots Q_N$ of the bilinear fields
while the $\Lambda^{2M}$ of the diagonal $SU(M)$ becomes
$\prod_\ell\Lambda_\ell^{2M}$, hence
the Seiberg--Witten curve
\bea
Y^2\ &=&\ \left[T(X)\right]^2\ -\ 4\Lambda^{2MN} \\
&=&\ \prod_{i=1}^M\left(X\,-\,V^N e^{Na\varphi_i}\right)^2\ -\ 4\Lambda^{2MN}
\nonumber
\eea

We may eliminate the $V^N$ parameter from this formula by rescaling other variables.
Let
\bea
x&=&\frac{X}{V^N},\quad
y\ =\ \frac{Y}{V^{MN}},\nonumber \\
r_i&=&\frac{R_i}{V^N}\ =\ \exp(Na\varphi_i),\nonumber \\
t(x)&=&\frac{T(X)}{V^{MN}}\ =\ \prod_{i=1}^M(x-r_i),\nonumber \\
\lambda&=&\frac{\Lambda^M}{V^M} \,;
\eea
then the rescaled Seiberg--Witten curve of the quiver becomes simply
\be
y^2\ =\ [t(x)]^2\ -\ 4\lambda^{2N} .
\label{SWCurve}
\ee
Thanks to the bound~(\ref{Vbound}), $|\lambda|\le 1$;
generically $|\lambda|<1$ and hence $\lambda^{2N}\ll 1$ in the large quiver limit
$N\to\infty$.
Consequently, the $2M$ branching points of the hyperelliptic curve~(\ref{SWCurve})
come in close pairs
\be
y\ =\ 0,\quad x\ =\ r_{i,\pm}\ =\ r_i\ \pm\ \Delta_i
\label{BranchPoints}
\ee
where
\be
\Delta_i\ =\ \frac{2\lambda^N}{t'(x=r_i)}\
\ll\ \left|r_i-r_j\right|\ \forall j\neq i
\label{Deltas}
\ee
{\em provided} all the eigenvalues $\phi_i=\Re\varphi$ are distinct.
That is, we focus on generic points of the Coulomb branch moduli space where
all the non-abelian symmetries are broken.
To avoid finite-size effects, we assume
\be
\forall\, i\neq j\quad|\phi_i-\phi_j|\ \gg\ \frac{1}{Na} ,
\ee
hence
\be
r_1\ \ll\ r_2\ \ll\ r_3\ \ll\ \cdots\ \ll\ r_M
\ee
(assuming the eigenvalues are ordered according to eq.~(\ref{Ordering}))
and therefore
\bea
t'(x=r_i) &=&
\prod_{j\neq i}(r_i-r_j)\ \approx\ (r_i)^{i-1}\times\prod_{j=i+1}^M(-r_j)\
\nonumber \\
&=& \frac{\pm1}{r_i}\times\prod_{j=i+1}^M\frac{r_i}{r_j}\ \gg\ \frac{1}{r_i}\,,
\eea
where the last equality follows from
$\sum_{j=1}^M\varphi_j=0\Rightarrow \prod_{j=1}^M r_j=1$.
Consequently,
\be
\Delta_i\ \approx\ \pm\ 2 r_i \lambda^N\times\prod_{j<i}\frac{r_j}{r_i}\
\ll r_i\lambda^N\ \ll\ r_i
\label{DeltaFormula}
\ee
and therefore indeed $\Delta_i\ll (r_i-r_j)\ \forall j\neq i$.

In the Appendix to this article, we show that
a hyperelliptic Seiberg--Witten curve whose branching points come in closed
pairs as in eq.~(\ref{BranchPoints}) yields weak gauge couplings.
Specifically, the holomorphic coupling matrix $\tau_{ij}$ is given by
\bea
\tau_{i\neq j} &=&
\frac{i}{2\pi}\,\log\frac{r_i r_j}{(r_i-r_j)^2} ,\\
\tau_{i=j} &=&
\frac{i}{2\pi}\,\log\frac{4r_i^2}{\Delta_i^2} .
\eea
For the problem at hand,
\bea
\nonumber
\frac{r_i r_j}{(r_i-r_j)^2}\
\approx\ \min\left(\frac{r_i}{r_j},\frac{r_j}{r_i}\right)\
=\ \exp\left(\pm Na(\varphi_i-\varphi_j)\right) ,
\eea
hence for $i\neq j$
\be
\tau_{ij}\ =\ \frac{i Na}{2\pi}\times\cases{
        (\varphi_i-\varphi_j) & for $i<j$,\cr
        (\varphi_j-\varphi_i) & for $j<i$.\cr }
\ee
At the same time,
\bea
\nonumber
\frac{4r_i^2}{\Delta_i^2}\ =\ \lambda^{-2N}\times
\prod_{k<i}\exp\left(2Na (\varphi_i-\varphi_k)\right)
\eea
and hence
\be
\tau_{ii}\ =\ \frac{i Na}{2\pi}\times
\left[ H\,+\,2\smash{\sum_{k<i}}\vphantom{\sum}(\varphi_i-\varphi_k)\right]\
\ee
where
\be
H\ =\ \frac{1}{a}\,\log\frac{1}{\lambda^2}\
=\ \frac{1}{a}\,\log\frac{V^{2M}}{\Lambda^{2M}}
\label{Hdef}
\ee

Note that (in the large quiver limit $N\to\infty$) the entire gauge coupling
matrix $\tau_{ij}$ is proportional to  the total length $L=Na$ of the deconstructed
fifth dimension, in perfect agreement with the Kaluza--Klein reduction.
Furthermore, by reversing the KK reduction~(\ref{GFourFive}) we immediately
obtain the 5D gauge couplings of the deconstructed theory,
\bea
\left[\frac{8\pi^2}{g_5^2}\right]_{ij} &
=& \delta_{ij}\times \left[ \Re H\,
        +\,2\smash{\sum_{k<i}}\vphantom{\sum}(\phi_i-\phi_k) \right]\
    -\ \left|\phi_i-\phi_j\right|
\nonumber \\
&=& \delta_{ij}\times \left[ \Re H\,
        +\,M\phi_i\,+\smash{\sum_{k=1}^M}\vphantom{\sum}|\phi_i-\phi_k|\right]\
    -\ \left|\phi_i-\phi_j\right| .
\label{DeconstructedG}
\eea
As promised, these couplings are {\em exactly} as in eq.~(\ref{SUMGaugeCouplings})
for the special case of Chern--Simons level $\kcs=M$,
provided we identify the tree-level 5D
inverse gauge coupling $h$ with $\Re H$,
\be
h\,\equiv\,\left[\frac{8\pi^2}{g_5^2(SU(M)}\right]^{\rm tree}\
=\ \Re H\ =\ \frac{1}{a}\,\log\left|\frac{V}{\Lambda}\right|^{2M} .
\label{TreeCoupling}
\ee
Thanks to the bound~(\ref{Vbound}), $h\ge 0$ throughout the entire
moduli space of the quiver and there are no ``flop'' transitions
to other phases of the 5D theory.
In other words, {\bf we have deconstructed the coupling parameter space of the
quantum 5D SYM theory as the positive real semi-axis $\bbbr^+$,
in perfect agreement with the M--theory results 
for $\kcs=M$}.

{}From the 5D SYM point of view, deconstructing the fifth dimension serves
as a UV completion of the theory in a radically different way that the M--theory
embeddings and the $(p,q)$ webs discussed in section~2.
Yet for all their differences, we end up with exactly the same parameter space
of the theory --- $h\ge 0$ only, no flop transitions --- which strongly suggests
that this parameter space is the inherent property of the 5D SYM regardless of the
UV completion one might want to use to work it out.

Note that the $\bbbr^+$ parameter space space without any flops
is peculiar to the maximal Chern--Simons level $\kcs=M$ of the 5D theory.
As we saw in the previous section,
a 5D SYM at a lower Chern--Simons level $\kcs<M$ ---
or rather its UV completion via IIB $(p,q)$ 5--brane webs
--- does have a flop transition at $h=0$
from the ordinary SYM phase at $h>0$ to something more exotic at $h<0$.
Again, this turns out to be an inherent property on the 5D theory, with
exactly the same phases and flop transitions in the deconstructed
theory as in the M--theory based completions.

Unfortunately, deconstructing the 5D theories with $\kcs<M$ requires more
complicated quivers than~(\ref{QuiverDiagram}),
so the authors prefer to present this issue in a separate follow-up article
\cite{IKtwo}.
We will show in detail how to deconstruct the 5D SQCD --- with or without flavors
--- for all allowed Chern--Simons numbers,
but our main conclusion can be stated right now:
{\em The deconstructed 5D SQCD has precisely the same {\bf 5D} phases and flop
transitions as the M--theory based UV completions of the same 5D theory.}
We emphasize the {\em 5D} phases here because some quiver theories also have
additional, strongly-coupled phases --- completely outside the 5D phase spaces ---
where the 4D gauge couplings are stuck at strong values, and the
deconstruction does not seem to work.
We suspect that such strongly-coupled 4D phases are related to superconformal
theories in 5D, but further research is needed before we can be more definite
on this issue.
But the ordinary, weakly-coupled 4D phases of the deconstructed theory are
precisely as in the M-based completions of the same 5D theory.

\subsection{Deconstructing Chern Simons}
\centerline{In collaboration with Edoardo Di~Napoli}

It is well known that
in a 5D gauge theory with 8 unbroken supercharges, the moduli dependence
of the gauge couplings is related to the topological Chern--Simons
couplings of the gauge fields, {\it cf.}\ eq.~(\ref{FiveDGauge}).
But it isn't known yet whether this relation survives breaking of 4 out of 8
supercharges in the process of dimensional deconstruction.
For example,
the quiver~(\ref{QuiverDiagram}) we used to deconstruct the 5D $SU(M)$ SYM
theory yielded gauge couplings corresponding to the Chern--Simons level
$\kcs=M$, and the $\bbbr^+$ geometry of the parameter space of the theory
gives further evidence for this identification.
But because of the partial SUSY breakdown, we cannot be sure about
the actual Chern--Simons interactions of between the 5D
gauge fields of the deconstructed theory, and the only way to be sure is to
derive them directly from the quiver theory.

In this subsection, we shall see that the deconstructed theory indeed has
Chern--Simons action
\be
S_{\rm CS}\ =\ \kcs\times\int\limits_{\bbbr^5}\!\!\wcs
\label{CSAction}
\ee
with the correct coefficient $\kcs=M$.
To stress the tree-level nature of these interactions from the 5D point of view,
we shall assume un-broken $SU(M)$ symmetry in 5D and look for the non-abelian
Chern--Simons 5--form
\be
\wcs\ =\ \frac{i}{24\pi^2}\,
\tr\left(A\wedge F\wedge F\,-\,\textstyle{i\over2} A\wedge A\wedge A\wedge F\,
-\,\textstyle{1\over 10}A\wedge A\wedge A\wedge A\wedge A\right).
\label{NACS}
\ee
Thus, we assume $\vev\phi=0$ and no Wilson lines,
or in quiver terms,
\be
\vev{Q_\ell}\,=\,V U_\ell,\quad
\mbox{unitary}\ U_\ell=(U_\ell^\dagger)^{-1},\quad
U_1 U_2\cdots U_N\,=\,1.
\label{UnbrokenSUM}
\ee
Likewise, in order to avoid unwarranted extended-SUSY based assumptions we shall work
in the component-field formalism where $\psi_\ell$ are the ${\cal N}=1$ fermionic
superpartners of the $Q_\ell$ scalars and $\lambda_\ell$ are the ${\cal N}=1$
gaugino fields of the $SU(M)_\ell$.
The Yukawa couplings of these fermions follow from the unbroken SUSY of the quiver
theory: In $M\times M$ matrix notations,
\be
{\cal L}_Y\
=\sum_\ell\tr\left(-ig Q_\ell^\dagger\lambda_\ell\psi_\ell\
+\ ig\psi_\ell\lambda_{\ell+1}Q_\ell ^\dagger\right)\ +\ \rm H.c.
\label{Yukawas}
\ee
Given the scalar VEVs~(\ref{UnbrokenSUM}), these Yukawa couplings produce 4D fermionic
mass terms with simple fixed eigenvalues but not-so-simple $U_\ell$ dependent
mass eigenstates: Let
\be
\tilde\psi_\ell\
=\ (U_1U_2\cdots U_{\ell-1})^\dagger\,\psi_\ell (U_1U_2\cdots U_{\ell-1}U_\ell) ,\qquad
\tilde\lambda_\ell\
=\ (U_1U_2\cdots U_{\ell-1})^\dagger\,\lambda_\ell (U_1U_2\cdots U_{\ell-1}) ,
\label{ChiralRot}
\ee
then the fermionic mass terms become
\be
{\cal L}^{\rm fermion}_{\rm masses}\ =\ \left\{
igV\sum_\ell\tr(-\tilde\lambda_\ell\tilde\psi_\ell+\tilde\psi_\ell\tilde\lambda_{\ell+1})\
=\, \sum_\ell i\tilde\psi_\ell\frac{\tilde\lambda_{\ell+1}-\tilde\lambda_\ell}{a}
\right\}\ +\ \mathrm{H.c.}
\ee
which obviously deconstructs the fifth component of the 5D fermion's kinetic energy term.
Indeed, combining $\tilde\psi_\ell$ with $\bar{\tilde\lambda_\ell}$ into a single 4--component
5D Dirac spinor $\Psi_\ell$, we have
\be
\sum_\ell i\tilde\psi_\ell\frac{\tilde\lambda_{\ell+1}-\tilde\lambda_\ell}{a}\
+\ \mathrm{H.c.}\
=\,\sum_\ell i\overline\Psi_\ell\gamma^5\frac{\Psi_{\ell+1}-\Psi_\ell}{a}\
\to\ \int\!\!dx^5\,i\overline\Psi\gamma^5\partial_5\Psi .
\ee

Note $U_\ell$ dependent chiral rotation~(\ref{ChiralRot}) involved in converting the original
fermionic fields $\psi_\ell$ and $\lambda_\ell$ of the quiver into the components $\tilde\psi_\ell$
and $\tilde\lambda_\ell$ (or rather $\bar{\tilde\lambda_\ell}$) of the 5D fermion $\Psi_\ell$.
Unless all the $U_\ell$ matrices are constant with respect to 4 continuous dimensions of spacetime,
the transform~(\ref{ChiralRot}) has a non-trivial functional determinant, which gives rise to the
Wess--Zumino terms for the $U_\ell(x)$ at the one-loop level of the perturbation theory.

By way of analogy, consider the Wess--Zumino terms on the ordinary QCD with $n_c>2$ colors and
$n_f>2$ flavors.
The spontaneous breakdown $SU(n_f)_L\times SU(n_f)_R\to SU(n_f)V$ of the chiral symmetry of the
theory gives rise to an effective non-linear sigma model of the matrix field $U(x)\in SU(n_f)$.
It also gives rise to a chiral transform of the original fermionic fields of the theory
--- the so-called {\it current} quark and antiquark fields  ---
into the eigenstates of the mass matrix known as the {\it constituent} quarks.
In terms of left-handed quarks $\psi$ and antiquarks $\chi$, the transform works as
\be
\psi'(x)\,=\,\psi(x) \times W(x)U(x),\qquad \chi'(x)\,=\,\chi(x)\times W(x)
\label{ChiralQCD}
\ee
for some arbitrary vector-like flavor symmetry $W(x)\in SU(n_f)_V$,
but the functional determinant  depends only on the chiral rotation $U(x)$.
The logarithm of this functional determinant is the Wess--Zumino action term for the pseudoscalar meson
fields comprising the $U$ matrix,
\be
S_{\rm WZ}\ =\ \log\Det\left(\psi(x)\to\psi(x)U(x)\right)\
=\ n_c\times \int\limits_{\bbbr^4}\!\wzf[U]
\ee
where the $n_c$ factor reflect $n_c$ identical multiplets of the chiral flavor symmetry suffering
the same chiral rotation $U(x)$.
The basic Wess--Zumino 4--form $\wzf$ is ugly and lacks manifest $SU(n_f)\times SU(n_f)$
symmetry, but its formal exterior derivative is nice and symmetric
\be
d\wzf[U]\ =\ \frac{i}{240\pi^2}\tr\left( J\wedge J\wedge J\wedge J\wedge J\right)\quad
\mathrm{where}\ J_\mu\,=\,-i(\partial_\mu U) U^\dagger.
\label{WZFF}
\ee
Therefore, the Wess--Zumino term is usually written as the 5D integral
\be
S_{\rm WZ}\ =\ n_c\int\limits_{{\cal B}^5}\!d\wzf[U]
\ee
over a solid ball ${\cal B}^5$ whose surface $S^4$ is topologically identified with the Euclidean
spacetime of the fermionic path integral.
That is, we introduce an auxiliary fifth dimension $t$ of finite range $0\le t\le1$ and extrapolate
\bea
U(x^1,x^2,x^3,x^4) &\mapsto& U(x^1,x^2,x^3,x^4,t)\in SU(n_f), \\
&& U(x^1,x^2,x^3,x^4,t=1)\, =\, U(x^1,x^2,x^3,x^4), \nonumber \\
&& U(x^1,x^2,x^3,x^4,t=0)\equiv 1. \nonumber
\eea
Consequently, $t$ acts as a radial coordinate of a 5--ball ${\cal B}^5$ while the
Euclidean spacetime coordinates $(x^1,x^2,x^3,x^4)$ act as its angular coordinates%
\footnote{We assume a well-defined limit $U(x\to\infty)$ as $x$ goes to infinity
        in any direction in the Euclidean space.
        Topologically, this makes the Euclidean space a sphere $S^4$.};
$J_\mu$ ($\mu=1,2,3,4,t$) is a matrix-valued 1--form on this 5--ball,
and $d\wzf$ of eq.~(\ref{WZFF}) is a 5--form.

In QCD, spontaneous breakdown of the chiral symmetry is a highly non-perturbative
dynamical effect.
However, once it happens, the Wess--Zumino term for the meson fields comprising $U(x)$
arises as a one-loop anomaly of the chiral transformation of the chiral fields.
After this transform, one may leave the constituent quark fields as a part of the effective
low-energy theory or integrate them out altogether:  One way or the other, the Wess--Zumino
terms remain a part of the meson field action.

In the quiver theory, the $[SU(M)]^N$ symmetry is spontaneously broken down (or rather
Higgsed down) at the tree level rather than dynamically,
but the effect on the fermionic fields of the theory is rather similar:
The original fermions $\psi_\ell$ and $\lambda_\ell$ of the quiver suffer a $U_\ell$ dependent
chiral transform~(\ref{ChiralRot}) into mass eigenstates which deconstruct the 5D fermion $\Psi$.
Chirally speaking, the transform~(\ref{ChiralRot}) combines a complicated but vector-like
transform (analogous to the $W(x)$ matrix in eq.~(\ref{ChiralQCD})) with a simple but chiral
transform
\be
\psi_\ell(x)\to \psi_\ell(x) U_\ell(x),\qquad \lambda_\ell(x)\to \lambda_\ell(x),
\ee
whose functional determinant gives rise to the QCD-like Wess--Zumino action
\be
S_{\rm WZ}\ =\ \sum_\ell\kwz\int\limits_{{\cal B}^5}\!d\Omega_{\rm WZ}(U_\ell)
\label{QuiverWZ}
\ee
whose coefficient follows from the multiplicity of the ``quark'' fields $\psi_\ell$:
As far as the $U_\ell$ is concerned, $\psi_\ell$ comprises $M$ ``colors'' undergoing
similar chiral transforms, hence
\be
\kwz\ =\ M.
\label{KWZM}
\ee

{}From the 5D point of view, the Wess--Zumino terms~(\ref{QuiverWZ}) deconstruct the
5D Chern--Simons action~(\ref{CSAction}) \cite{SkibaSmith,Napoli}.
To see how this works, we need to consider not just five but six spacetime dimensions:
The four continuous dimensions $x^1,\ldots x^4$ of the quiver theory, the
deconstructed fifth dimension $x^5=a\ell$ discretized via quiver label $\ell$,
and finally the auxiliary dimension $t$ of the Wess--Zumino action.
Please note that although both the $x^5$ and the $t$ are often called ``the fifth dimension,''
they are quite distinct and should not be confused with each other.
In eq.~(\ref{QuiverWZ}), the 5--ball ${\cal B}^5$ spans $(x^1,x^2,x^3,x^4,t)$
while the $x^5$ dimension is discretized as $\ell$.

We must also modify eq.~(\ref{WZFF}) for the Wess--Zumino 5--form to account for the gauging
of the $SU(M)_\ell\times SU(M)_{\ell+1}$ ``flavor'' symmetry of each $U_\ell$.
Altogether, the Wess--Zumino action comes to
\be
S_{\rm WZ}\ =\,\sum_\ell\frac{i\kwz}{240\pi^2}\int\limits_{{\cal B}^5}
\tr\left\{ \matrix{ 
J\wedge J\wedge J\wedge J\wedge J\
    -\ 5i J\wedge J\wedge J\wedge (F+H) \cr
{}+ 5J\wedge\bigl( 2F\wedge F\,+\,2H\wedge H\,
        +\,F\wedge H\,+\,H\wedge F\bigr)\cr
}\right\}_\ell
\label{WZAction}
\ee
where
\bea
J_\ell &=& -i\,DU_\ell^{}\,U_\ell^\dagger\
    =\ -i\,dU_\ell^{}\,U_\ell^\dagger\ +\ A_\ell\ -\ U_\ell^{} A_{\ell+1} U_\ell^\dagger, \\
D\,J_\ell &=& iJ_\ell\wedge J_\ell\ +\ F_\ell\ -\ H_\ell ,\nonumber \\
F_\ell &=& dA_\ell\,+\,iA_\ell\wedge A_\ell,\nonumber \\
H_\ell &=& U_\ell^{} F_{\ell+1} U_\ell^\dagger , \nonumber
\eea
and the gauge-anomaly terms contained in each $d\wcs[U_\ell]$ cancel out of the $\sum_\ell$.
The action (\ref{WZAction}) is valid for arbitrary $SU(M)$ matrices $U_\ell(x)$,
but for the deconstruction purposes we are interested in the continuous $x^5$
limit $a\to 0$ where
\be
U_\ell\ =\ 1\ +\ ia A_5\ +\ O(a^2)\quad \mathrm{at}\ x^5=a\ell ,
\ee
{\it cf.}~eq.~(\ref{QuiverVEV}).
In this limit $H_\ell=F_\ell+O(a)$ while
\be
J_{\ell,\mu}\
=\ a\left( \partial_\mu A_5\,-\,\partial_5 A_\mu\,+\,i[A_\mu,A_5] \right)\
+\ O(a^2)\ \equiv\ a F_{\mu 5}\ +\ O(a^2),
\ee
or in the 5D 1--form notations, $J_\ell=aF^5_\ell+O(a^2)$.
Consequently, to the leading order in $a$,
\bea
S_{\rm WZ}\ &=&
\sum_\ell\frac{i\kwz}{240\pi^2} \int\limits_{{\cal B}^5}\tr\left\{
30a\, F^5_\ell\wedge F_\ell\wedge F_\ell\ +\ O(a^2) \right\} \\
&\longrightarrow&
\frac{i\kwz}{8\pi^2}\int\limits_{S^1}\!dx^5\!\int\limits_{{\cal B}^5}
        \tr\Bigl( F^5\wedge F\wedge F\Bigr)_{\rm 5D} \nonumber \\
&=& \frac{i\kwz}{24\pi^2}\! \int\limits_{S^1\times{\cal B}^5}
        \tr\Bigl(F\wedge F\wedge F \Bigr)_{\rm 6D} . \nonumber
\eea
But in six dimensions
\be
\frac{i}{24\pi^2}\tr\left( F\wedge F\wedge F\right)\ =\ d\wcs
\ee
is the total derivative of the Chern--Simons 5--form~(\ref{NACS}),
hence in the continuum limit
\be
\setbox0=\hbox{$\,a\to 0\,$}
S_{\rm WZ}\ \mathop{\hbox to \wd0{\rightarrowfill}}\limits_{\box0}\
\kwz\!\int\limits_{S^1\times{\cal B}^5}\!\!d\wcs\
=\ \kwz\!\int\limits_{S^1\times S^4}\!\!\wcs
\label{CSAD}
\ee
where the deconstructed fifth dimension $x^5$ spans the circle $S^1$, the continuous
spacetime dimensions $(x^1,x^2,x^3,x^4)$ are topologically compactified to $S^4$ and the
auxiliary dimension $t$ has disappeared from the final formula.
Apart from topological subtleties of the deconstructed 4D spacetime being $S^1\times S^4$
instead of $S^5$ or $\bbbr^5$, the action~(\ref{CSAD}) is precisely the
Chern--Simons action (\ref{CSAction}) we wanted to deconstruct, with precisely the right
coefficient
\be
\kcs\ =\ \kwz\ =\ M.
\ee

This somewhat lengthy exercise in ``topological deconstruction'' verifies $\kcs=M$
for the quiver~(\ref{QuiverDiagram}), but it also helps us answer a deeper question,
namely {\em how does one deconstruct a 5D SYM with $\kcs\neq M$?}
Clearly, the Wess--Zumino / Chern--Simons connection is rather rigid and does not
allow any deviations from $\kcs=\kwz$.
On the other hand, we can change the Wess--Zumino level of the quiver if and only if
we change the fermionic fields or the chiral transforms they suffer while being
packaged into the 5D fermions.
Without breaking the ${\cal N}=1$ SUSY of the quiver theory, we cannot tinker with
$\psi_\ell$ and the $\lambda_\ell$ fermions, but we may add more fermions
to the theory.
In the follow-up article \cite{IKtwo} we shall use this approach do deconstruct
the 5D SQCD --- with or without flavors --- at any allowed Chern--Simons level
\be
|\kcs|\ \le\ n_c\ -\ \frac{n_f}{2}\,,\qquad \kcs\ \in\ \bbbz\,+\,\frac{n_f}{2}\,.
\ee

\section*{Acknowledgments}
The authors would like to thank Jacob Sonnenschein, Yaron Oz and Edoardo Di~Napoli
for valuable discussions.
This research  was supported by the US National 
Science Foundation (grants NSF--DMS/00--74329 and PHY--0071512),
the Robert A.~Welsh foundation and the
US--Israel Bi--National Science Foundation.

\newpage
\section*{Appendix: Hyperelliptic Seiberg--Witten Curves}
Consider an abelian rank $r$ Coulomb branch of a generic 4D supersymmetric
gauge theory.
The holomorphic gauge couplings of the abelian fields form an $r\times r$
symmetric matrix $\tau_{ij}$ defined modulo $Sp(r,\bbbz)$ symmetry group
of electric-magnetic dualities.
The duality-independent representation of the coupling matrix is given by
a hyperelliptic Seiberg--Witten curve of genus $g=r$, generally of the
form
\be
y^2\ =\ \prod_{i=1}^{2r+2}(x-b_i)
\label{GenericSW}
\ee
where the branching points $b_i$ are distinct for generic values of the theory's
moduli but collide when some charged particles become massless.

The first homology $H^{(1)}(\bbbz)$ of the curve (\ref{GenericSW}) has
a symplectic product --- the intersection number.
Choosing a particular symplectic basis of 1--cycles $\alpha_1,\ldots,\alpha_r,
\beta_1,\ldots,\beta_r$,
\be
\alpha_i\cap\beta_j=\delta_{ij}\,,\qquad
\alpha\cap\alpha\equiv\beta\cap\beta\equiv0,
\label{Symplexis}
\ee
corresponds to choosing a basis of the electric and the magnetic charges and hence
of the electric and the magnetic fields.
In this basis, the coupling matrix $\tau_{ij}$ relates the periods of
holomorphic 1--forms of the curve,
\bea
\label{GenericPeriods}
\oint\limits_{\beta_i}\!\Omega &
=&\sum_j\tau_{ij}\times\oint\limits_{\beta_i}\!\Omega ,\\
\Omega &
=& \frac{dx}{y}\times \mathop{\rm Polynomial}(x)\
\mbox{of degree}\ (r-1).
\eea

In this Appendix we evaluate the couplings of hyperelliptic curves whose branching
points come in very close pairs,
\bea
y^2 &=& \prod_{i=1}^M(x-r_{i,+})(x-r_{i,-}) ,
\label{TheSW} \\
\mbox{\vrule height 15 pt width 0pt}
r_{i,\pm} &=& r_i\pm\Delta_i,\qquad
|\Delta_i|\ \ll\ |r_i-r_j|\ \forall\,j\neq i .
\label{ClosePairs}
\eea
Geometrically, such a curve looks like a pair of complex spheres
connected by $M$ very thin tubes,
\be
\vcenter{\openup 1\jot \ialign{%
        $\displaystyle{y\ \approx\ #}$\hfil\quad &
        for $x$ #{}\hfil \cr
        \pm\prod_{i=1}^M(x-r_i) &
        away from all $r_i$\cr
        \omit\hfil\vrule depth 10pt width 0pt &
        \omit\quad {\it i.e.}, $\forall i: (x-r_i)\gg\Delta_i$,\hfil\cr
        \pm\sqrt{(x-r_i)^2-\Delta_i^2}\times\prod_{j\neq i}(x-r_j) &
        near $r_i$,\cr
        }}
\label{CurveGeometry}
\ee
which gives us a natural choice of the ``electric'' $\alpha$ cycles of the curve:
$\alpha_i$ runs around the tube around $r_i$.
Equivalently, we let the square-root branch cuts of the curve (\ref{TheSW})
connect each branching point $r_{i,\pm}$ to its closest neighbor $r_{i,\mp}$
and let the cycle $\alpha_i$ run around the cut between the $r_{i,-}$ and the $r_{i,+}$.
Notice that homologically
\be
\alpha_1\,+\,\alpha_2\,+\,\cdots\,+\,\alpha_M\ =\ 0
\label{AlphaRedundency}
\ee
and only $r=M-1$ of these $\alpha_i$ cycles are independent.
This redundancy matches  the redundancy of the abelian gauge
field basis $A^\mu_i$ we use in the main body of this article:
For the Coulomb branch of a 5D $SU(M)$ theory, the $A^\mu_i$ are superpartners
of the scalar eigenvalues $\phi_i$, hence
\be
M\ \mbox{vector fields}\
A^\mu_1,A^\mu_2,\ldots,A^\mu_M\ \mbox{constrained to satisfy}\
A^\mu_1\,+\,A^\mu_2\,+\,\cdots\,+\,A^\mu_M\ =\ 0 ,
\ee
and we follow the same convention for the 4D quiver.
Consequently, the corresponding $M\times M$ matrix $\tau_{ij}$ of the gauge couplings
is redundant modulo
\be
\tau_{ij}\ \to\ \tau'_{ij}\ =\ \tau_{ij}\,+\,c_i\,+\,c_j\quad
\forall (c_1,c_2,\ldots,c_M).
\label{TauRedundency}
\ee

The ``magnetic'' $\beta$ cycles of the SW curve (\ref{TheSW}) run through
(rather than around) the connecting tubes at $r_i$; two tubes
are involved in each $\beta$ cycle.
This gives us a very large set of magnetic cycles to choose from:
For any $j\neq k$, we can set up the branch cuts of the curve such that the
$r_{j,+}$ branching point is connected to the $r_{k,-}$; the cycle surrounding
this cut (and no other cuts) we call $\beta_{jk}$.
Naturally, only $M-1$ of these magnetic cycles are independent while the rest follow
from the linear relations
\be
\beta_{ij}\ +\ \beta_{jk}\ =\ \beta_{ik}\ +\ \alpha_j .
\ee
Furthermore, the $\beta_{jk}$ cycles are not exactly dual to the electric
cycles $\alpha_i$:
Instead of the symplectic intersections (\ref{Symplexis}), we have
\be
\alpha\cap\alpha\ \equiv\ 0,\qquad \beta\cap\beta\ \equiv\ 0,\qquad
\mbox{but}\quad \alpha_i\cap\beta_{jk}\ =\ \delta_{ij}\,-\,\delta_{ik}\,.
\label{IntersectAB}
\ee
Consequently, the periods of holomorphic 1--forms of the SW curve are
related to the gauge couplings according to
\be
\oint\limits_{\beta_{jk}}\!\Omega\
= \sum_{i=1}^M\left( \tau_{ij}\,-\,\tau_{ik}\,+\,\mbox{an integer}\right)\times
\oint\limits_{\alpha_i}\!\Omega\
\label{PeriodsTaus}
\ee

As usual, $\Omega=(dx/y)\times {}$a polynomial of degree $M-2$ or less.
To simplify the period calculations, we let
\be
\Omega\ =\ \frac{dx}{y}\times\sum_{i=1}^M C_i\prod_{j\neq i}(x-r_j),\qquad
\sum_{i=1}^M C_i\,=\,0.
\label{PolynomialParams}
\ee
hence in light of the curve geometry (\ref{CurveGeometry}),
\be
\vcenter{\openup 1\jot \ialign{%
        $\displaystyle{\Omega\ \approx\ #}$\hfil\quad &
        for $x$ #{}\hfil \cr
        \pm\sum_{i=1}^M\frac{C_i\,dx}{x-r_i} &
        away from all $r_i$,\cr
        \pm\left[\frac{C_i\,dx}{\sqrt{(x-r_i)^2-\Delta_i^2}}\
            +\,\sum_{j\neq i}\frac{C_j\,dx}{x-r_j}\right] &
        near $r_i$,\cr
        \pm\sum_{i=1}^M\frac{C_i\,dx}{\sqrt{(x-r_i)^2-\Delta_i^2}} &
        everywhere. \cr
        }}
\label{OmegaForm}
\ee
In terms of the $x$ coordinate, an electric cycle $\alpha_i$ is basically a small
circle surrounding $r_i\pm\Delta_i$, hence the corresponding period of the $\Omega$
is simply
\be
\oint\limits_{\alpha_i}\!\Omega\
\approx \oint\limits_{\alpha_i}\!\frac{C_i\,dx}{\sqrt{(x-r_i)^2-\Delta_i^2}}\
+\,\sum_{j\neq i}\oint\limits_{\alpha_i}\!\frac{C_i\,dx}{x-r_i}\
=\ 2\pi i C_i\ +\ 0.
\ee
For a magnetic cycle $\beta_{jk}$, the integration contour runs from the $r_{j,+}$
to the $r_{k,-}$ on one Riemann sheet, then comes back on the other sheet where
all the square roots take opposite signs.
Hence, the magnetic period
\bea
\oint\limits_{\beta_{jk}}\!\Omega &
\approx& -2\int\limits_{r_{j,+}}^{r_{k,-}}\sum_i
        \frac{C_i\,dx}{\sqrt{(x-r_i)^2-\Delta_i^2}}\
        =\ -2\left.\sum_i C_i\mathop{\rm ar}\sinh\frac{x-r_i}{\Delta_i}\right|_{r_{j,+}}^{r_{k,-}}
    \hbox{\vrule depth 30pt width 0pt} \\
&\approx& 2\left[ C_j\log\frac{\Delta_j}{2}\,+\sum_{i\neq j}C_i\log(r_j-r_i)\right]\
        -\ 2\left[ C_k\log\frac{\Delta_k}{2}\,+\sum_{i\neq k}C_i\log(r_j-r_k)\right] .
\nonumber
\eea
Comparing these periods with eqs.~(\ref{PeriodsTaus}), we immediately arrive at
the gauge coupling matrix
\be
\tau_{ij}\ =\ \frac{2}{2\pi i}\left\{
    \vcenter{\openup 1\jot \ialign{%
        $\displaystyle{\log #}$\quad\hfil & for #\hfil\cr
        (r_i-r_j) &  $i\neq j$,\cr
        \frac{\Delta_i}{2} &  $i=j$,\cr
  }}\right\}\
+\ O\left(\frac{\Delta^2_{i,j}}{(r_i-r_j)^2}\right) ,
\ee
or equivalently --- thanks to the redundancy (\ref{TauRedundency}),
\bea
\tau_{i\neq j} &=& \frac{i}{2\pi}\log\frac{r_i\,r_j}{(r_i-r_j)^2},
\nonumber \\
\tau_{ii} &=& \frac{i}{2\pi}\log\frac{4r_i^2}{\Delta_i^2} .
\label{SWTaus}
\eea

Physically, the limit of very close pairs of branching points
corresponds to the weak gauge couplings.
Indeed, eqs.~(\ref{SWTaus}) show that in the $\Delta_i\to 0$ limit,
\be
g_i^2\ \approx\ \frac{4\pi^2}{\log\left|r_i/\Delta_i\right|}\
+\ O\left(\frac{1}{\log^2}\right)\ \to 0.
\ee
To get a more accurate picture, let us focus on the coupling of a particular
abelian gauge field.
For example, let $A^\mu_{ij}=A^\mu_i-A^\mu_i$; this field has coupling
\bea
\tau(A^\mu_{ij}) &=& \tau_{ii}\,+\,\tau_{jj}\,-\,2\tau_{ij}
\nonumber \\
&=& \frac{i}{2\pi}\log\frac{16(r_i-r_j)^4}{\Delta_i^2\Delta_j^2}\
        +\ O(\left(\frac{\Delta^2}{(r_i-r_j)^2}\right)
\nonumber \\
&=& \frac{2i}{2\pi}\log(4\chi_{ij})\ +\ O(1/\chi_{ij})
\label{ijTau}
\eea
where
\be
\chi_{ij}\ =\ 4\frac{(r_{i,+}-r_{j,-})(r_{i,-}-r_{j,+})}
        {(r_{i,+}-r_{i,-})(r_{j,+}-r_{j,-})}\
\approx\ \frac{(r_i-r_j)^2}{\Delta_i\Delta_j}
\ee
is the $SL(2,\bbbc)$ invariant cross-ratio of the 4 branching points
$r_{i,\pm}$ and $r_{j,\pm}$.
When the branching points come in closed pairs~(\ref{ClosePairs}),
all the $\chi_{ij}$ invariants are large, and all the gauge couplings are weak.

Note that the physical gauge couplings due to a Seiberg--Witten curve are
invariant under the $SL(2,\bbbc)$ transforms of the complex $x$ plane.
Hence, eqs.~(\ref{ijTau}) remain valid as long as all $\chi_{ij}\gg1$,
even if some of the $r_{i,\pm}$ pairs are not close.
The close-pairs condition~(\ref{ClosePairs}) is sufficient to assure that all
the physical gauge couplings are weak and calculable according to our formulae,
but it is stronger than necessary.

In our next article \cite{IKtwo} we will show that deconstructing 5D SQCD with
flavors or with Chern--Simons levels $C<M$ leads to Seiberg--Witten curves
which do not always satisfy the close-pairs condition~(\ref{ClosePairs}).
When a pair $r_{i,\pm}$ splits apart but all the cross-invariants $\chi_{ij}$
are large, the theory undergoes a flop transition in the large quiver limit.
But when some of the cross-invariants become small, the low-energy 4D couplings
become large and the deconstruction fails altogether.

\newpage


\begin{thebibliography}{99}

\bibitem{KLMVW}
A. Klemm, W. Lerche, P. Mayr, C. Vafa, N. Warner, ``Self-Dual Strings and $N=2$
Supersymmetric Field Theory,'' Nucl. Phys. {\bf B477} (1996) 746-766, {\tt hep-th/9604034}.

\bibitem{witten97}
E. Witten, ``Solutions Of Four-Dimensional Field Theories Via 
M Theory,'' Nucl. Phys. {\bf B500} (1997) 3-42, {\tt hep-th/9703166}.

\bibitem{KKV}
S. Katz, A. Klemm, C. Vafa, ``Geometric Engineering of Quantum Field Theories,''
Nucl. Phys. {\bf B497} 173-195, {\tt hep-th/9609239}.

\bibitem{seiberg}
Nathan Seiberg, {\it``Five Dimensional SUSY Field Theories, Non-trivial
Fixed Points and String Dynamics,''\/ \tt hep-th/9608111}.

\bibitem{witten96}
E. Witten, ``Physical Interpretation of Certain Strong Coupling Singularities,''
Mod. Phys. Lett. {\bf A11} (1996) 2649-2654, {\tt hep-th/9609159}.

\bibitem{DKV}
M.~R.\ Douglas, S.~Katz and C.~Vafa, {\it``Small Instantons, del Pezzo
Surfaces and Type~$I'$ Theory,'' \sl Nucl. Phys. \bf B497} (1997) 155--172,
{\tt hep-th/9609071}.

\bibitem{MS}
D.~R.\ Morrison and N.~Seiberg, {\it``Extremal Transitions
and Five Dimensional Supersymmetric Field Theories,''
\sl Nucl. Phys. \bf B483} (1997) 229, {\tt hep-th/9609070}.

\bibitem{IMS}
K.~Intriligator, D.~R.\ Morrison and N.~Seiberg, {\it``Five-Dimensional
Supersymmetric Gauge Theories and Degeneration of Calabi-Yau Spaces,''
\sl Nucl. Phys. \bf B497} (1997) 56, {\tt hep-th/9702198}.

\bibitem{AHK}
O.~Aharony, A.~Hanany and B.~Kol,
{\it``Webs of $(p,q)$ 5--Branes, Five Dimensional
Field Theories and Grid Diagrams,'' \sl JHEP \bf 9801} (1998) 002,
{\tt hep-th/9710116}.

\bibitem{LV}
N.~C.\ Leung and C.~Vafa, {\it``Branes and Toric Geometry,''
\sl Adv. Theor. Math. Phys. \bf 2} (1998) 91--118, {\tt hep-th/9711013}.

\bibitem{KR2}
B.~Kol and J.~Rahmfeld, {\it``BPS Spectrum of 5 Dimensional Field Theories,
$(p,q)$ Webs and Curve Counting,'' \sl JHEP \bf 9808} (1998) 006, {\tt hep-th/9801067}.

\bibitem{PhasesW}
Edward Witten, {\it``Phase Transitions in M--Theory and F--Theory,''
\sl Nucl. Phys. \bf B471} (1996) 195--216, {\tt hep-th/9603150}.

\bibitem{ACG}
N.~Arkani--Hamed, A.~G.\ Cohen and H.~Georgi,
{\it``(De)Constructing Dimensions,''
\sl Phys. Rev. Lett. \bf 86} (2001) 4757, {\tt hep-th/0104005}.

\bibitem{IKtwo}
Amer Iqbal, Vadim S.\ Kaplunovsky and Edoardo Di~Napoli,
{\it``Quantum Deconstruction of 5D SQCD
and its Moduli Space for General $n_c$, $n_f$ and $\kcs$,''\/} in preparation.

\bibitem{csaki}
C.~Csaki, J.~Erlich, C.~Grojean and G.~Kribs,
{\it``4D Constructions of Supersymmetric Extra
Dimensions and Gaugino Mediation,'' \sl Phys. Rev. \bf D65} (2002) 015003,
{\tt hep-th/0106044}.

\bibitem{csaki2}
C.~Csaki, J.~Erlich, V.~V.\ Khoze, E.~Poppitz, Y.~Shadmi and Y.~Shirman,
{\it``Exact Results in 5D from Instantons and Deconstruction,''
\sl Phys. Rev. \bf D65} (2002) 085033, {\tt hep-th/0110188}.

\bibitem{seibergq}
Nathan Seiberg, {\it``Exact Results on the Space of Vacua of Four Dimensional
SUSY Gauge Theories,''\/ \tt hep-th/9402044}.

\bibitem{paper1}
C.~Csaki, J.~Erlich, D.~Freedman and W.~Skiba,
{\it``"N=1 Supersymmetric Product Group
Theories in the Coulomb Phase,''\/ \tt hep-th/9704067}.

\bibitem{wittenlinear}
Edward Witten, {\it``Phases of $N=2$ Theories in Two Dimensions,''
\sl Nucl. Phys. \bf B403} (1993) 159--222, {\tt hep-th/9301042}.

\bibitem{rodriguez}
J.~M.\ Rodr\'\i guez, {\it``Vacuum Structure of Supersymmetric $SU(N)^K$ Theories,''\/}
Ph.~D.\ Thesis (1997), UMI--98--25065--mc (microfiche).

\bibitem{georgi}
S.~Chang and H.~Georgi, {\it``Quantum Modified Mooses''\/ \tt hep-th/0209038}.

\bibitem{SkibaSmith}
W.~Skiba and D.~Smith, {\it``Localized Fermions and Anomaly Inflow via Deconstruction,''
\sl Phys. Lett. \bf D65} (2002) 095, {\tt hep-ph/0201056}.

\bibitem{Napoli}
E.\ Di~Napoli, unpublished work.

\bibitem{MorrisonG}
David Morrison, {\it``Geometric Aspects of Mirror Symmetry,''\/ \tt math.AG/0007090}.

\end{thebibliography}
\end{document}